\documentclass[a4paper,twocolumn,aps,nofootinbib,prd]{revtex4}
\usepackage[latin9]{inputenc}
\usepackage[unicode = true,
 bookmarks = true,bookmarksnumbered = false,bookmarksopen = false,
 breaklinks = false,pdfborder = {0 0 0},backref = false,colorlinks = true]
 {hyperref}
\usepackage{breakurl}

\makeatletter

\@ifundefined{textcolor}{}
{%
 \definecolor{BLACK}{gray}{0}
 \definecolor{WHITE}{gray}{1}
 \definecolor{RED}{rgb}{1,0,0}
 \definecolor{GREEN}{rgb}{0,1,0}
 \definecolor{BLUE}{rgb}{0,0,1}
 \definecolor{CYAN}{cmyk}{1,0,0,0}
 \definecolor{MAGENTA}{cmyk}{0,1,0,0}
 \definecolor{YELLOW}{cmyk}{0,0,1,0}
}

\usepackage{epsfig}
\usepackage{graphicx}
\usepackage{epsf}
\usepackage{amssymb,amsmath}
\usepackage[usenames]{color}
\usepackage[usenames]{xcolor}
\usepackage{amssymb}
\usepackage{times}
\usepackage{mathrsfs}
\usepackage{hyperref}
\hypersetup{
  colorlinks=true,        
  linkcolor=blue,         
  citecolor=cyan,         
}
\newcommand{\cf}{cf.~}
\newcommand{\ie}{i.e.,~}
\newcommand{\eg}{e.g.,~}

\definecolor{green}{rgb}{0.15,0.7,0.15}

\makeatother

\begin{document}

\title{Effect of intense magnetic fields on reduced-MHD evolution in
  $\sqrt{s_{\rm NN}}$ =  200 GeV Au+Au collisions}

\author{Victor Roy$^{1}$, Shi Pu$^{2}$, Luciano Rezzolla$^{3,4}$, Dirk H.\ Rischke$^{3,{5}}$}

\affiliation{$^{1}$National Institute of Science Education and Research,
HBNI, 752050 Odisha, India. }
\affiliation{$^{2}$Department of Physics, The University of Tokyo, 
  7-3-1 Hongo, Bunkyo-ku, Tokyo 113-0033, Japan}
\affiliation{$^{3}$Institute for Theoretical Physics, Goethe University, 
Max-von-Laue-Str.\ 1, D-60438 Frankfurt am Main, Germany}
\affiliation{$^{4}$Frankfurt Institute for Advanced Studies,
Ruth-Moufang-Str.\ 1, D-60438 Frankfurt am Main, Germany}
\affiliation{{$^{5}$Department of
Modern Physics, University of Science and Technology of China, Hefei,
Anhui 230026, China}}

\begin{abstract}
We investigate the effect of large magnetic fields on the $2+1$
dimensional reduced-magnetohydrodynamical expansion of hot and dense
nuclear matter produced in $\sqrt{s_{\rm NN}}$ = 200 GeV Au+Au
collisions. For the sake of simplicity, we consider the case where the
magnetic field points in the direction perpendicular to the reaction
plane. We also consider this field to be external, with energy density
parametrized as a two-dimensional Gaussian. The width of the Gaussian
along the directions orthogonal to the {beam axis} varies {with the
  centrality of the collision}. The dependence of the magnetic field on
proper time ($\tau$) for the case of zero electrical conductivity of the
QGP is parametrized following Ref.~\cite{Deng:2012pc}, and for finite
electrical conductivity following Ref.~\cite{Tuchin:2013apa}. We solve
the equations of motion of ideal hydrodynamics for such an external
magnetic field. For collisions with {non-zero impact parameter} we
observe considerable changes in the evolution of the momentum
eccentricities of the fireball when comparing the case when the magnetic
field decays in a conducting QGP medium and when no magnetic field is
present. The elliptic-flow coefficient $v_2$ of $\pi^{-}$ is shown to
{increase} in the presence of an external magnetic field and the
increment in $v_2$ is found to depend on the evolution and the initial
magnitude of the magnetic field.
\end{abstract}
\maketitle

\section{Introduction}
\label{sec:introduction}

Two positively charged heavy nuclei produce ultra-intense magnetic fields
in collider experiments at the Relativistic Heavy Ion Collider~(RHIC) and
at the Large Hadron Collider~(LHC), e.g.\ $B \sim\,10^{18} -10^{19}\,{\rm
  G}$ for $\sqrt{s_{\rm NN}} = $ 200 GeV Au+Au collisions. The intensity
of the magnetic field in the transverse plane grows approximately
linearly with the center-of-mass energy ($\sqrt{s_{{\rm NN}}}$)
\cite{Bzdak:2011yy,Deng:2012pc, Tuchin:2013apa} {[see also recent studies
    including non-zero chiral conductivity \cite{Tuchin:2014iua,
      Li:2016tel}]}. The corresponding electric field in the transverse
plane also becomes very large since it is enhanced by a Lorentz
factor. Such intense electric and magnetic fields are believed to have a
strong impact on the dynamics of high-energy heavy-ion collisions. For
example, a strong magnetic field may induce energy loss of fast quarks
and charged leptons via synchrotron radiation \cite{Tuchin:2010vs}, or
{may} enhance dilepton and photon
production~\cite{Tuchin:2010gx,Basar:2012bp}. There are several other
interesting phenomena related to the presence of ultra-intense magnetic
fields in heavy-ion collisions. For example, in the case of an imbalance
in the number of left- vs.\ right-handed fermions, a charge current is
induced in the Quark-Gluon Plasma~(QGP), leading to the separation of
electrical charges, which is known as the ``chiral magnetic effect''
(CME) \cite{Kharzeev:2007jp}. Within a 3+1 dimensional anomalous
hydrodynamics calculation, Ref.~\cite{Hirono:2014oda} showed that the CME
could be seen in azimuthal correlations of charged hadrons. Along with
the CME, it was also theoretically predicted that massless fermions with
the same charge but different chirality will be separated, known as
``chiral separation effect'' (CSE). A connection between these effects
and the Berry phase in condensed-matter {systems} was pointed out in
Refs.~\cite {Son:2012wh, Son:2012zy, Stephanov:2012ki, Gao:2012ix,
  Chen:2012ca,Hidaka:2016yjf}, and some nonlinear chiral transport
phenomena {were studied} in Refs.~\cite{Pu:2014fva, Chen:2016xtg,
  Ebihara:2017suq,Gorbar:2016qfh}. Within the statistical
hadron-resonance gas model of Ref.~\cite{Bhattacharyya:2015pra},
significant changes of hadron multiplicities were observed in the
presence of a strong magnetic field. Finally, the possibility of a change
in the quark-hadron phase transition line in the QCD phase diagram under
the combined influence of external magnetic field and local vortices was
explored in Ref.~\cite{McInnes:2016dwk}; we refer the reader to the recent
reviews \cite{Kharzeev:2013ffa, Bzdak:2012ia, Kharzeev:2015kna,
  Tuchin:2013ie,Huang:2015oca}, where more details can be found.
 
Relativistic dissipative hydrodynamics has so far been successfully
applied to explain the experimentally measured flow harmonics in
heavy-ion collisions. The success of hydrodynamics implies that a QGP
with small shear-viscosity to entropy-density ratio is formed in Au+Au
collisions at top RHIC energies within a short time interval $\sim 0.2 -
0.6\,{\rm fm}$ \cite{Shen:2011zc,Romatschke, Heinz, Bozek:2012qs,
  Roy:2012jb, Heinz:2011kt, Niemi:2012ry, Schenke:2011bn}. The system is
  close to local equilibrium, thus the initial geometry of the collision
  has a strong influence on the final momentum anisotropy. However, the
possible effect of a magnetic field on the hydrodynamical evolution has
so far not been studied extensively, except for some simplified cases
\cite{Gursoy:2014aka,Zakharov:2014dia} and most recently using some
approximate form of the equations of relativistic magnetohydrodynamics
(MHD)~\cite{Pang:2016yuh,Das:2017qfi}, or employing a 3+1-dimensional
partonic cascade BAMPS (Boltzmann Approach to Multi-Parton Scatterings)
\cite{Greif2017}.

In a parallel analytical approach, in Refs.~\cite{Roy:2015kma,
  Pu:2016ayh,Pu:2016bxy} solutions of the ideal-MHD equations were found
in simplified geometries. More specifically, for Au+Au collisions at
$\sqrt{s_{\rm NN}} = $ 200 GeV, the electromagnetic field energy was
shown to be comparable to the initial energy density of the QGP in
Ref.~\cite{Roy:2015coa}. In a recent work \cite{Mohapatra:2011ku}, it was
argued that a magnetic field of magnitude $eB \sim m_{\pi}^2 \sim 10^{18}
-10^{19}\,{\rm G}$, with $m_{\pi}$ the pion mass, can induce a large
azimuthal anisotropy of the produced particles. In
Refs.~\cite{Tuchin:2011jw,Critelli:2014kra} it was also shown that the
magnitude of the shear viscosity extracted from the experimental data is
underestimated when ignoring the magnetic field. On the other hand,
Ref.~\cite{Pang:2016yuh} has found that the elliptic flow is reduced in
the presence of a magnetic field when one considers a
temperature-dependent magnetic susceptibility of the QGP. Similarly, in
Ref.~\cite{Voronyuk:2011jd} magnetic fields were found to have only a
very small impact on the flow harmonics within the Parton Hadron String
model.

Here we will study the $2+1$ dimensional expansion of matter with
vanishing magnetization in terms of the dynamics of a perfect fluid
\cite{Rezzolla_book:2013} in the presence of an external magnetic
field. We refer to this approach as to ``reduced MHD'' and we note that
this is not a self-consistent solution of the full set of MHD equations,
since we only use a parametrized form for the evolution of the magnetic
field and do not solve Maxwell's equations together with the conservation
equations of energy and momentum. For the sake of simplicity, we also
assume that the electrical conductivity is infinite (\ie the ideal-MHD
limit), since this allows to eliminate the electric field in favor of the
magnetic field (see below). Contrasting our approach with the one of
Ref.~\cite{Inghirami:2016iru}, it is useful to remark that they are quite
complementary. In fact, while in Ref.~\cite{Inghirami:2016iru} the full
set of ideal-MHD equations was employed, it was solved only for a
comparatively small value of the initial magnetic field and for a simple
ultrarelativistic equation of state (EOS). Here instead, we employ the
reduced-MHD formulation, but study the impact of varying the initial
magnetic field strength, adopting, furthermore, a realistic EOS.

We should also note that, in principle, one should then not use a
parametrized form for the magnetic-field evolution, because for a
perfectly conducting fluid one can show that the magnetic field follows
the evolution of the entropy density [the so-called ``frozen-flux''
  theorem, see Refs.~\cite{Roy:2015kma,Pu:2016ayh}]. Vice versa, using
some parametrized form of the magnetic field generally implies that the
electric conductivity is finite. Assuming a perfectly conducting fluid
under the influence of an external magnetic field still represents a
reasonable first approximation, which however calls for a future
improvement towards a self-consistent MHD solution, along the lines of
the work carried out in Ref.~\cite{Inghirami:2016iru}. We also assume
that the magnetic field only points into the $y$-direction. In
Ref.\ \cite{Roy:2015coa} this was shown to be a good approximation for
peripheral collisions. In a first approximation, we will also neglect 
\textcolor{black}{the magnetization of the QGP and} the
change in the EOS due to the magnetic field. We then investigate the
effect of the magnetic field on the fluid evolution and the momentum
anisotropy of charged particles on an event-averaged basis. The goal of
our study is to clarify how large the external magnetic field has to be
and how slowly it has to decay in order to make a sizable impact on the
momentum anisotropy of charged particles.

The paper is organized as follows: in Sec.~\ref{sec:mat_setup} we discuss
the mathematical formalism employed in our calculations, while the
numerical set-up is presented in Sec.~\ref{sec:num_setup}. Our results
are discussed in detail in Sec.~\ref{sec:results} and a summary is given
at the end in Sec.~\ref{sec:summary}. We use natural units $\hbar = c =
\epsilon_0 = \mu_0 = 1$, where $\epsilon_0$ and $\mu_0$ are the electric
permittivity and magnetic permeability in vacuum, respectively, and the
electric charge $e := \sqrt{4\pi\hbar c\alpha}\simeq 0.303$, where
$\alpha \simeq 1/{137}$ is the fine-structure constant. In these units
the quantity $eB$ has dimension $\rm GeV^{2}$. Throughout the paper the
components of four-tensors are indicated with Greek indices, whereas
three-vectors are denoted as boldface symbols. The metric tensor
in flat spacetime is $g^{\mu \nu} = {\rm diag}\, (+,-,-,-,)$.

\section{Magnetohydrodynamics}
\label{sec:mat_setup}

\textcolor{black}{We consider a system consisting of ``matter'',
  represented by a QGP with electric charge, and ``fields'', i.e.,
  electromagnetic fields which are created in the collision of heavy
  ions. The spacetime evolution of the coupled system of QGP and
  electromagnetic field is obtained by solving the equations of motion of
  MHD, i.e., energy-momentum conservation coupled to Maxwell's
  equations. In order to relate our work to that of others, we first
  discuss the MHD equations of non-dissipative, polarized, and magnetized
  fluids in general \cite{degroot,Israel:1978up,Kovtun:2016lfw}, and then
  specialize to the case of a perfectly conducting, non-dissipative
  fluid.}

\subsection{MHD of non-dissipative, polarized, and magnetized fluids}
The energy-momentum conservation equation reads
\begin{equation}
\label{eq:dmu_Tmunu}
\partial_{\nu}T^{\mu\nu}  =  0\;, 
\end{equation}
with $T^{\mu\nu}$ being the total energy-momentum tensor.  The latter can
be decomposed into a matter part, $T^{\mu \nu}_{{\rm mat}}$, and a field
part, $T^{\mu \nu}_{{\rm field}}$, such that
\begin{equation}
T^{\mu \nu} = T^{\mu \nu}_{{\rm mat}} + T^{\mu \nu}_{{\rm field}}\;,
\end{equation}
but this decomposition is not unique. Following Israel
\cite{Israel:1978up}, for a non-dissipative, polarized, and magnetized
fluid we define (note that our convention for the metric tensor differs
from that of Israel \cite{Israel:1978up} by an overall sign)
\begin{eqnarray}
T^{\mu \nu}_{{\rm mat}} & := & (\varepsilon + p) u^\mu u^\nu - p\, g^{\mu \nu} - \Pi^\mu u^\nu\;, \label{eq:Tmat} \\
T^{\mu \nu}_{{\rm field}} & := & F^\mu_{\hspace*{0.2cm} \alpha} H^{\alpha \nu} 
+ \frac{1}{4} g^{\mu \nu} F_{\alpha \beta} F^{\alpha \beta}\;, \label{eq:Tfield}
\end{eqnarray}
where $\varepsilon$ and $p$ are energy density and pressure of the fluid,
respectively, and $u^{\mu} :=\gamma (1, \boldsymbol{v})$ is the
four-velocity of the fluid in an arbitrary frame (in our context we
choose the center-of-momentum (CM) frame of the heavy-ion collision),
where the fluid moves with three-velocity $\boldsymbol{v}$; $\gamma
:=(1-\boldsymbol{v}^2)^{-1/2}$ is the Lorentz factor\footnote{Note that
  hereafter we will indicate spatial three-vectors with a bold face, \ie
  $\boldsymbol{V} = \vec{\mathbf{V}}$}. Introducing antisymmetrization of
a rank-2 tensor $A^{\mu \nu}$ via the notation $A^{[\mu \nu]} :=
\frac{1}{2} (A^{\mu \nu} - A^{\nu \mu})$, the auxiliary vector $\Pi^\mu$
in Eq.\ (\ref{eq:Tmat}) is defined as \cite{Israel:1978up}
\begin{equation} \label{eq:auxPi}
\Pi^\mu := 2 u_\lambda F^{[\mu}_{\hspace*{0.3cm} \nu} M^{\lambda] \nu} \;, 
\end{equation}
where
\begin{equation} \label{eq:Faraday}
F^{\mu \nu} = E^\mu u^\nu - E^\nu u^\mu + \epsilon^{\mu \nu \alpha \beta} u_\alpha B_\beta
\end{equation}
is the Faraday tensor. Here, $\epsilon^{\mu \nu \alpha \beta}$ is the
completely antisymmetric four-tensor, $\epsilon^{0123} = \sqrt{\det|g|}$
with $g_{\mu\nu}$ being the metric tensor, $E^\mu := F^{\mu \nu} u_\nu$
is the electric field and $B^\mu := \frac{1}{2}\epsilon^{\mu \nu \alpha
  \beta} u_\nu F_{\alpha\beta}$ the magnetic induction field, both
measured in a frame comoving with the fluid. Note that by definition
$E^\mu$ and $B^\mu$ are orthogonal to $u^\mu$, \ie $E^\mu u_\mu = B^\mu u_\mu
=0$.  Also, both $E^\mu$ and $B^\mu$ are space-like vectors, \ie $0> E^\mu
E_\mu$ and $0> B^\mu B_\mu =: - B^2$. 

The in-medium Faraday tensor in Eq.\ (\ref{eq:Tfield}) is defined as
$H^{\mu \nu} := F^{\mu \nu} - M^{\mu \nu}$, where
\begin{equation} \label{eq:polarization}
M^{\mu \nu} = -P^\mu u^\nu + P^\nu u^\mu + \epsilon^{\mu \nu \alpha \beta} u_\alpha M_\beta\;,
\end{equation}
is the polarization tensor, also appearing in Eq.\ (\ref{eq:auxPi}), with
the polarization vector $P^\mu := - M^{\mu \nu} u_\nu$ and the
magnetization vector $M^\mu:= \frac{1}{2} \epsilon^{\mu \nu \alpha \beta}
u_\nu M_{\alpha \beta}$.  Note that also $P^\mu$ and $M^\mu$ are
orthogonal to $u_\mu$, \ie $P^\mu u_\mu = M^\mu u_\mu =0$, as well as
being space-like, \ie $0> P^\mu P_\mu$, $0> M^\mu M_\mu$. Hereafter, we
will assume that $P^\mu = \chi_E E^\mu$ and $M^\mu = \chi_B B^\mu$, which
is characteristic for matter with a linear response to electromagnetic
fields.

Inserting Eqs.\ (\ref{eq:Faraday}) and (\ref{eq:polarization}) into
Eq.\ (\ref{eq:auxPi}) yields
\begin{equation}
\Pi^\mu = \epsilon^{\mu \nu \alpha
  \beta}u_\nu (M_\alpha E_\beta - P_\alpha B_\beta)\;,   
\end{equation}
so that in the comoving frame
\begin{equation}
\Pi^0 =0\;, \quad {\rm and} \quad \boldsymbol{\Pi} = \boldsymbol{P}
\times \boldsymbol{B}- \boldsymbol{M} \times \boldsymbol{E}\;.
\end{equation}

Note that neither $T^{\mu \nu}_{\rm mat}$ nor $T^{\mu
    \nu}_{\rm field}$ are by themselves symmetric, but their sum is,
  $T^{\mu \nu} = T^{\nu \mu}$. To see this, compute their antisymmetric
  parts $T^{[\mu \nu]}_{\rm mat} = - \Pi^{[\mu} u^{ \nu]}$ and $T^{[\mu
      \nu]}_{\rm field} = - F^{[\mu}_{\hspace*{0.3cm} \alpha} H^{\nu]
    \alpha} \equiv F^{[\mu}_{\hspace*{0.3cm} \alpha} M^{\nu] \alpha}$ and
  use the identity [see Eq.\ (6.24) of Ref.\ \cite{Israel:1978up}]
\begin{equation} \label{eq:identity}
\Pi^{[\mu} u^{\nu]} = F^{[\mu}_{\hspace*{0.3cm} \alpha} M^{\nu] \alpha}\;,
\end{equation}
which can be readily proven using Eqs.\ (\ref{eq:Faraday}) and
(\ref{eq:polarization}), together with the assumption that the response
of the matter to electromagnetic fields is linear.

A decomposition of the energy-momentum tensor where each term is
symmetric by itself reads \cite{Israel:1978up}
\begin{equation}
T^{\mu \nu} = T^{\mu \nu}_{\rm sym} + T^{\mu \nu}_{\rm free\, field}\;,
\end{equation}
with the symmetric ''free'' energy-momentum tensor of the electromagnetic
field
\begin{equation}
T^{\mu \nu}_{\rm free \, field} := F^\mu_{\hspace*{0.2cm} \alpha} F^{\alpha \nu} 
+ \frac{1}{4} g^{\mu \nu} F_{\alpha \beta} F^{\alpha \beta}\;,
\end{equation}
and the symmetric ``matter'' energy-momentum tensor
\begin{eqnarray}
T^{\mu \nu}_{\rm sym} &:=& T^{\mu \nu}_{\rm mat} +
  F^{\mu}_{\hspace*{0.2cm} \alpha} M^{\nu \alpha}
\nonumber \\ 
& = & (\varepsilon+p)u^\mu u^\nu - p g^{\mu \nu} - \Pi^{(\mu} u^{\nu)} +
F^{(\mu}_{\hspace*{0.25cm} \alpha} M^{\nu) \alpha}\;,
\nonumber \\ 
\end{eqnarray}
where we used Eq.\ (\ref{eq:identity}) and introduced a symmetrized
rank-2 tensor via the notation $A^{(\mu \nu)} := \frac{1}{2} ( A^{\mu
  \nu} + A^{\nu \mu})$.

Note that the definitions of energy-momentum tensor $T^{\mu\nu}_{\rm
  mat}$ in Refs.~\cite{Pu:2016ayh, Huang:2011dc} do not contain the terms
proportional to the auxiliary vector $\Pi^\mu$. This is because for the
physical conditions encountered in relativistic heavy-ion collisions,
both electromagnetic susceptibilities $\chi_E$ and $\chi_B$, as well as
the ratio of the electromagnetic energy density to fluid energy density
are usually much smaller than unity \cite{Roy:2015coa}, so that $\Pi_\mu
/(\varepsilon+p) \sim \chi_{E,B} B^2/(\varepsilon+p) \ll 1$, and the
auxiliary vector $\Pi^\mu$ can be neglected as a first approximation.

Maxwell's equations in matter read
\begin{equation} \label{eq:Maxwell}
\partial_\mu H^{\mu \nu} = j^\nu\;, \;\;\; \partial_\mu \tilde{F}^{\mu \nu} =0\;,
\end{equation}
where $j^\nu := \rho u^\nu$ is the electric-charge four-current, with the
net electric charge density $\rho$, and $\tilde{F}^{\mu\nu} :=\frac{1}{2}
\epsilon^{\mu \nu \alpha \beta} F_{\alpha \beta}$ is the dual Faraday
tensor.  Using these equations, one can show that
\begin{equation}
\partial_\nu T^{\mu \nu}_{\rm field} = - F^{\mu \nu} j_\nu + \frac{1}{2} M_{\alpha \beta} \partial^\mu F^{\alpha \beta}\;.
\end{equation}
Moreover, using the Boltzmann equation, Israel \cite{Israel:1978up}
proved that
\begin{equation} \label{eq:evol_Tmat}
\partial_\nu T^{\mu \nu}_{\rm mat} = F^{\mu \nu} j_\nu - \frac{1}{2} M_{\alpha \beta} \partial^\mu F^{\alpha \beta}\;,
\end{equation}
so that the sum of both equations indeed gives total energy-momentum
conservation, Eq.\ (\ref{eq:dmu_Tmunu}).  This implies that the symmetric
``matter'' energy-momentum tensor obeys the equation
\begin{equation} \label{eq:evol_Tsym}
\partial_\nu T^{\mu \nu}_{\rm sym} = F^{\mu \nu} ( j_\nu + \partial_\lambda M^{\lambda}_{\hspace{0.2cm} \nu})\;.
\end{equation}

\textcolor{black}{
\subsection{Ideal MHD}
The electric current induced by an electric field is $j^\mu_{\rm ind} :=
\sigma E^\mu$, where $\sigma$ is the electric conductivity. Since for a
perfect conductor, $\sigma \rightarrow \infty$, we have to demand that
$E^\mu \rightarrow 0$, otherwise the induced current would be
infinite. This simplifies the equations of motion of MHD considerably,
because in this case also $P^\mu = \chi_E E^\mu \rightarrow 0$, which
eliminates the auxiliary vector $\Pi^{\mu}$ in Eq. (\ref{eq:auxPi}) from
the discussion. The ``matter'' energy-momentum tensor becomes that of a
non-dissipative fluid in the absence of fields,
\begin{equation}
T^{\mu \nu}_{{\rm mat}} \rightarrow (\varepsilon + p) u^\mu u^\nu - p\, g^{\mu \nu} \;, \label{eq:Tmat_ideal} 
\end{equation}
while the symmetric ``matter'' energy-momentum tensor assumes the form
given in Eq.\ (4) of Ref.\ \cite{Huang:2011dc},
\begin{equation}
T^{\mu \nu}_{\rm sym} \rightarrow (\varepsilon+p)u^\mu u^\nu - p g^{\mu \nu} 
+ F^{(\mu}_{\hspace*{0.25cm} \alpha} M^{\nu) \alpha}\;.
\end{equation}
For a linear response of matter to the magnetic induction field, $M^\mu =
\chi_B B^\mu$, the total energy-momentum tensor can then be brought into
the form} \cite{Pu:2016ayh, Gedalin:1995, Huang:2011dc,
  Giacomazzo:2005jy}\footnote{Note that this form of the energy-momentum
  tensor is different from the one normally used in general-relativistic
  formulations of the equations of MHD. In particular, in that notation
  $b^{\mu}$ are the contravariant components of the magnetic field in the
  frame comoving with the fluid; see Appendix A of
  Ref.~\cite{Roy:2015kma} for a more detailed discussion.}
\begin{align}
\label{eq:emt}
T^{\mu\nu} = & \left(\varepsilon+p-M B+B^{2}\right)u^{\mu}u^{\nu} \nonumber \\ 
- & \left(p-MB+\frac{1}{2}B^{2}\right)g^{\mu\nu}+ \left(MB-B^{2}\right)b^{\mu}b^{\nu}
\nonumber \\
= & \left[\varepsilon+p+B^2\left(1-\chi_B\right)
\right]u^{\mu}u^{\nu} \nonumber \\ -& \left[p+
\frac{1}{2}B^{2}\left(1-2\chi_B\right)\right]g^{\mu\nu}
+ B^2\left(1-\chi_B\right)b^{\mu}b^{\nu}\;,
\end{align}
where $M = \sqrt{-M^\mu M_\mu}$ and $b^{\mu} = B^{\mu}/B$.  Note that
because of $E^\mu = 0$, the electric field $\boldsymbol{\bar{E}}$ in the
CM frame can be eliminated in favor of the magnetic induction field
$\boldsymbol{\bar{B}}$ in the CM frame via $\boldsymbol{\bar{E}} = -
\boldsymbol{v} \times \boldsymbol{\bar{B}}$.  This implies $B^{2} =
\boldsymbol{\bar{B}}^2 (1-\boldsymbol{v}^2) + (\boldsymbol{v} \cdot
\boldsymbol{\bar{B}})^2$. Note also that the magnetization in the
comoving frame is actually defined as $M^\mu := \chi H^\mu = \chi
B^\mu/(1+\chi)$, where $H^\mu = B^\mu - M^\mu$ is the magnetic field in
the comoving frame and $\chi = \chi_B/(1-\chi_B)$ is the magnetic
susceptibility. If the latter is very small, then to first order $\chi
\simeq \chi_B$, and the magnetization can be approximated as $M^\mu
\simeq \chi B^\mu + \mathcal{O}(\chi^{2})$.  Since the magnetic
susceptibility $\chi \ll 1$ in the temperature range applicable for
heavy-ion collisions, \ie $\chi \lesssim 0.05$ for $eB \sim 0.2\,{\rm
  GeV}^2$ \cite{Bonati:2013lca}, we will set $M= 0$ in the actual
calculations.

\textcolor{black}{
\subsection{Reduced-MHD evolution}
As discussed above, a consistent MHD evolution would require to solve
Maxwell's equations (\ref{eq:Maxwell}) simultaneously with the
energy-momentum conservation equation (\ref{eq:dmu_Tmunu}).  In this
work, we do not attempt this rather formidable task, but restrict
ourselves to the so-called ``reduced-MHD'' set-up, where the magnetic
field evolution is prescribed from outside and only the energy-momentum
conservation equation is solved.}

The evolution of magnetic field considered here follows that of
Ref.~\cite{Tuchin:2013apa}. The physical picture is the following:
although the magnetic field produced at the time of collisions is large,
it also decays very quickly due to the high velocity of the
spectators. According to the Maxwell equation
$\nabla\times\boldsymbol{\bar{E}} = -\partial_{t}\boldsymbol{\bar{B}}$ ,
a time-varying magnetic field induces an electric field, which, in turn,
will produce an electric current $\boldsymbol{j}$ in the QGP medium that
depends on the conductivity and the displacement current in the
medium. This induced current will give rise to an induced magnetic field
in the same direction as the original magnetic field and hence the net
magnetic field is expected to decay more slowly than if the evolution
took place in vacuum.

\begin{figure}[h]
\vspace*{0.3cm}
\includegraphics[width=\columnwidth]{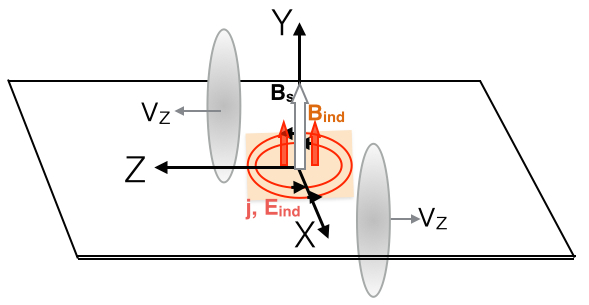}
\caption{Schematic diagram of a typical non-central heavy-ion collision
  and the corresponding electromagnetic fields in the reaction zone.}
\label{fig:Collision}
\end{figure}

The physical conditions just described above are shown schematically in
Fig.~\ref{fig:Collision}. The initial large but time-varying magnetic
field produced mostly due to the spectators is shown as
$\boldsymbol{B}_{\rm s}$, whereas the induced magnetic field is shown by
red arrows and denoted as $\boldsymbol{B}_{\rm ind}$. The induced
electric field in the reaction plane and the corresponding current
$\boldsymbol{j}$ are shown by the red circles. We remark that the
calculation of Ref.\ \cite{Tuchin:2013apa} assumes a constant electric
conductivity, but in our case the system evolves in space and time, so
that the electrical conductivity of the plasma should not be taken to be
constant but a function of temperature.

At this point, let us briefly comment on the treatment of the
conservation equations in Ref.~\cite{Pang:2016yuh}. A common feature to
our work is that the authors of Ref.~\cite{Pang:2016yuh} also assumed an
ideally conducting fluid, $E^\mu \rightarrow 0$. There are, however, two
important differences to our work: (1) the magnetization $M$ was assumed
to be non-zero and (2) the effect of the magnetic field $B$ in the
energy-momentum conservation equation was neglected.  In essence,
Ref.~\cite{Pang:2016yuh} just solved the evolution equation
(\ref{eq:evol_Tmat}) for the ``matter'' part of the energy-momentum
tensor under the assumption of a vanishing electric-charge four-current
$j^\mu = 0$, but for non-vanishing magnetization $M$. In this case, using
the relations
\begin{equation}
M^\nu= M b^\nu\;, \quad B^\nu = B b^\nu\;, \quad b^\nu b_\nu =-1\;,
\end{equation}
such that $b^\nu \partial^\mu b_\nu =0$, Eq.\ (\ref{eq:evol_Tmat})
then reads
\begin{equation} 
\label{eq:evol_Tmat_simple}
\partial_\nu T^{\mu \nu}_{\rm mat} = - M \partial^\mu B\;.
\end{equation}
Equation \eqref{eq:evol_Tmat_simple} differs by a sign from Eqs.\ (2) and
(3) of Ref.~\cite{Pang:2016yuh}. However, note that the EOS of state used
in the fluid evolution in Ref.~\cite{Pang:2016yuh} did not include the
effect from the magnetic field. As discussed in
Ref.\ \cite{Huang:2011dc}, in this case one needs to replace $\varepsilon
\rightarrow \varepsilon - MB$, $p \rightarrow p + MB$, such that the
right-hand side of Eq.\ (\ref{eq:evol_Tmat_simple}) is replaced by $+ B
\partial^\mu M$.  For a constant magnetic susceptibility $\chi$, this is
then equivalent to Eq.\ (4) of Ref.~\cite{Pang:2016yuh}. However, that
work used a temperature-dependent $\chi$, cf.\ their Eq.\ (5).

\textcolor{black}{
\subsection{2+1 dimensional geometry}}
We will assume a Bjorken-scaling expansion in the longitudinal direction,
so that, on account of boost invariance, we may restrict the discussion
to the $z=0$ plane, where for reasons of symmetry $u^z =0$. In this case,
it is advantageous to use Milne coordinates $\left(\tau,x,y,\eta\right)$,
where $\tau := \sqrt{t^2-z^2}$, $\eta := (1/2) \ln [ (t+z)/(t-z)]$, and the
metric tensor is given by $g^{\mu\nu} = {\rm
  diag}\left(1,-1,-1,-1/\tau^2\right)$. The energy-momentum conservation
equations \eqref{eq:dmu_Tmunu} then take the following form
%
{
\begin{widetext}
\begin{eqnarray}
\label{eq:EMConsT} 
 \partial_{\tau} \tilde{T}^{\tau\tau} +
 \partial_{x}\left(\tilde{T}^{\tau\tau} \tilde{v}^x \right) +
 \partial_{y}\left(\tilde{T}^{\tau\tau}\tilde{v}^y \right)  &=&  - p_B  + \tau f_B \tilde{B}^2 (b^\eta)^2
  \;,
\\
\label{eq:EMConsX}
 \partial_{\tau}\tilde{T}^{\tau x} +   \partial_{x}\left(\tilde{T}^{\tau x}  v^{x} \right) +
   \partial_{y}\left(\tilde{T}^{\tau x} v^{y} \right) 
   & = &   -   \partial_{x}\left[\tilde{p}_B - f_B \tilde{B}^2 b^x (b^{x}
   - b^{\tau} v^x)\right] + \partial_{y} \left[f_B \tilde{B}^2 b^x (b^{y}- b^{\tau} v^y)\right]
   \;, \\
\label{eq:EMConsY}
  \partial_{\tau} \tilde{T}^{\tau y} +    \partial_{x}\left(\tilde{T}^{\tau y}  v^{x} \right) 
  +   \partial_{y}\left(\tilde{T}^{\tau y} v^{y} \right)
     & = &  -  \partial_{y}\left[\tilde{p}_B - f_B\tilde{B}^2  b^y(b^{y}- b^{\tau}v^y)\right] 
     + \partial_{x}\left[f_B \tilde{B}^2 b^y (b^x -  b^{\tau}v^x)\right]
    \;,
\end{eqnarray}
\end{widetext}
where we have defined 
\begin{equation}
p_{B} := p- M B +\frac{B^2}{2}\;, \;\; \; f_B := 1 - \frac{M}{B}\;,
\end{equation}
as well as $\tilde T^{\mu\nu} := \tau T^{\mu\nu}$,
$\tilde{p}_B := \tau p_B$,
$\tilde{B}^2 := \tau B^2$, and
\begin{align}
\label{eq:vtildex}
\tilde{v}^x :=& \frac{T^{x\tau}}{T^{\tau \tau}} = \frac{w\gamma^2 v^x
  -f_B B^2 b^{x}b^{\tau} }{w\gamma^2-p_B-f_B 
  B^2 (b^{\tau})^2} \;,\\
\label{eq:vtildey}
\tilde{v}^y :=& \frac{T^{y \tau}}{T^{\tau \tau}} = \frac{w\gamma^2 v^y
  -f_B B^2 b^{y}b^{\tau} }{w\gamma^2-p_B-f_B 
   B^2 (b^{\tau})^2}\;,
\end{align}
with $w: = \varepsilon + p + f_B B^{2}$. Note that, at $\eta = 0$, $b^i -
b^\tau v^i = \bar{B}^i/(\gamma B)$. Note also that, at $\eta = 0$, $b^\eta
= \bar{B}^z/(\gamma B)$, which vanishes if the magnetic field
$\boldsymbol{\bar{B}}$ has no component in beam direction.

From Eq.~(\ref{eq:EMConsY}) it is clear that a magnetic field along the
$y$-direction decreases the total pressure 
However, since what drives the evolution of the fluid are the pressure
{\it gradients}, a constant magnetic field does not lead to a change of
the fluid acceleration. That said, and we will see below, the spatial
distribution of {the magnetic field} is such that also pressure gradients
are enhanced (reduced) along the $x$ ($y$)-axis,
respectively. Ultimately, this will result in an increase in the
momentum-space anisotropy of the fluid.

The set of equations (\ref{eq:EMConsT})--(\ref{eq:EMConsY}) is closed by
an EOS and we use the EOS indicated as ``s95p-PCE165-v0'' in
Refs.~\cite{Huovinen:2009yb,Shen:2010uy}, which is constructed from
lattice-QCD data at high temperature and a partially chemically
equilibrated hadron resonance gas at low temperature. From now on we will
refer to this as EOS-LHRG.  \textcolor{black}{Note that, for $M=0$,
  neither $\varepsilon$ nor $p$ change due to a non-vanishing
  magnetization energy density.}\\[0.5cm]

\section{Numerical setup}
\label{sec:num_setup}

We solve the conservation equations
(\ref{eq:EMConsT})--(\ref{eq:EMConsY}) {for $M =0$, \ie $f_B
  \equiv 1$,} by using an {appropriately} modified {(see below)} version
of {the} publicly available $2+1$ dimensional perfect
fluid dynamics code ``\texttt{AZHYDRO}'' \cite{Kolb:1999it, Roy:2011xt},
which uses the multidimensional flux-correcting algorithm \texttt{SHASTA}
to solve the energy-momentum conservation equations.

\begin{figure*}
\includegraphics[width=0.33\textwidth]{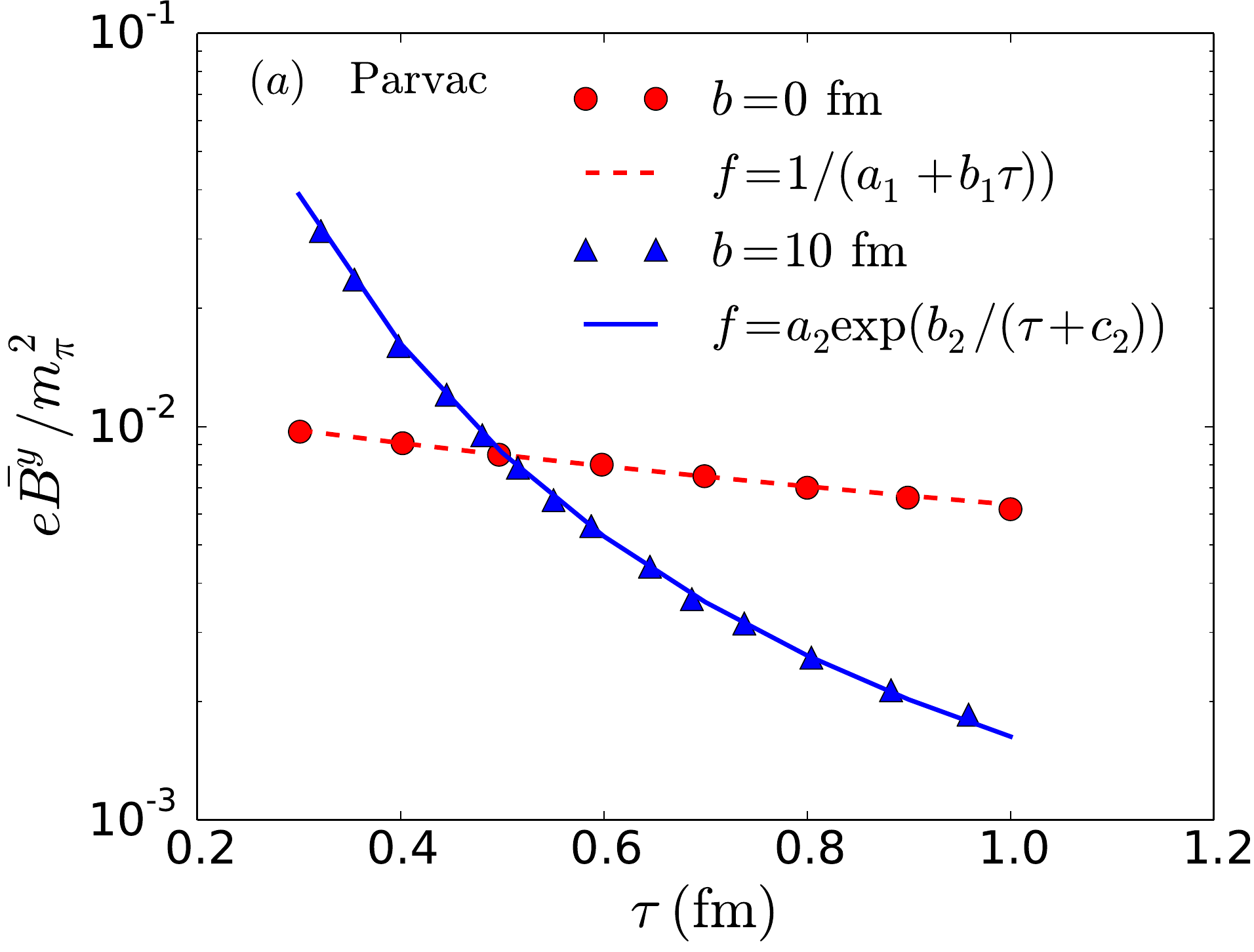}
\includegraphics[width=0.33\textwidth]{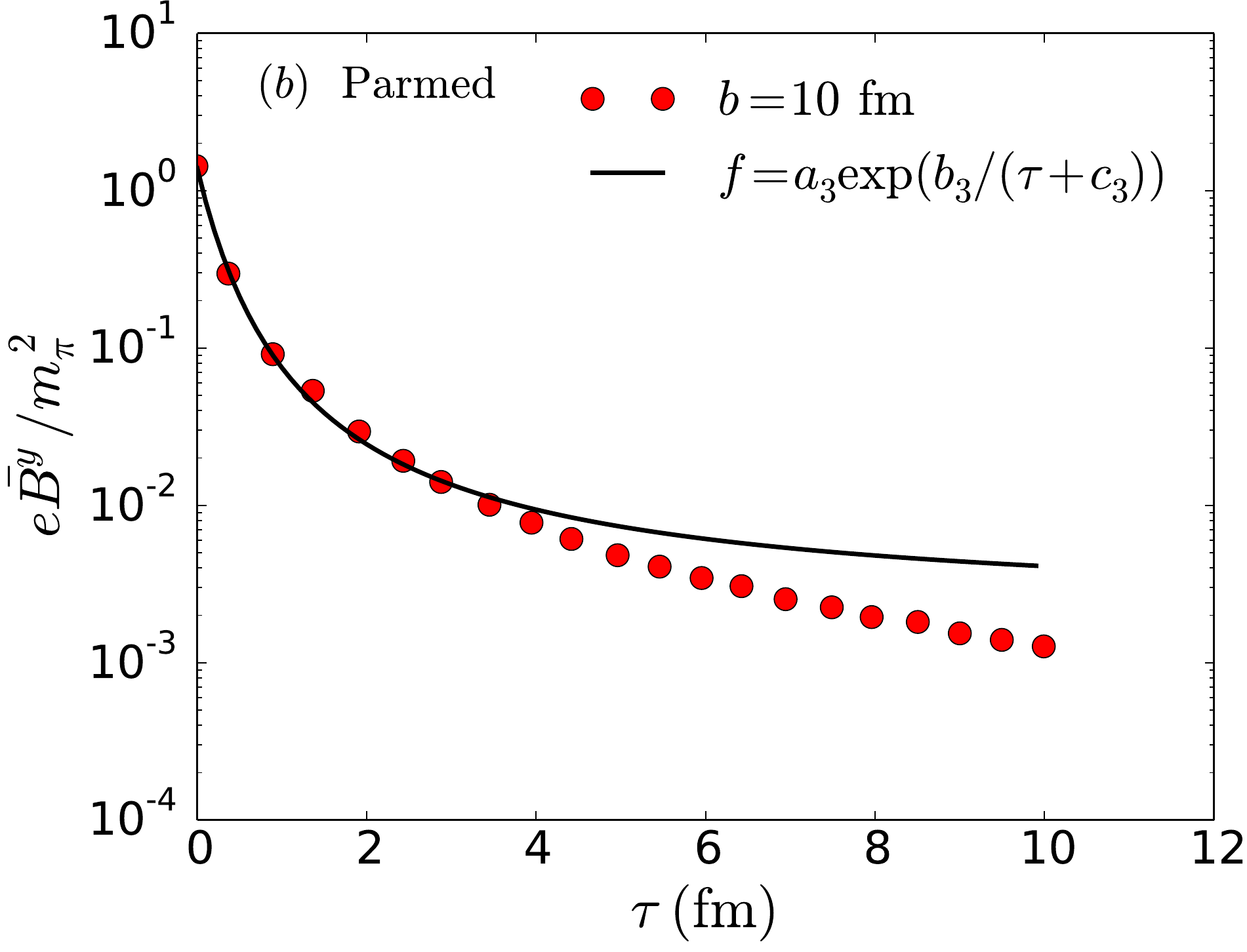}
\includegraphics[width=0.33\textwidth]{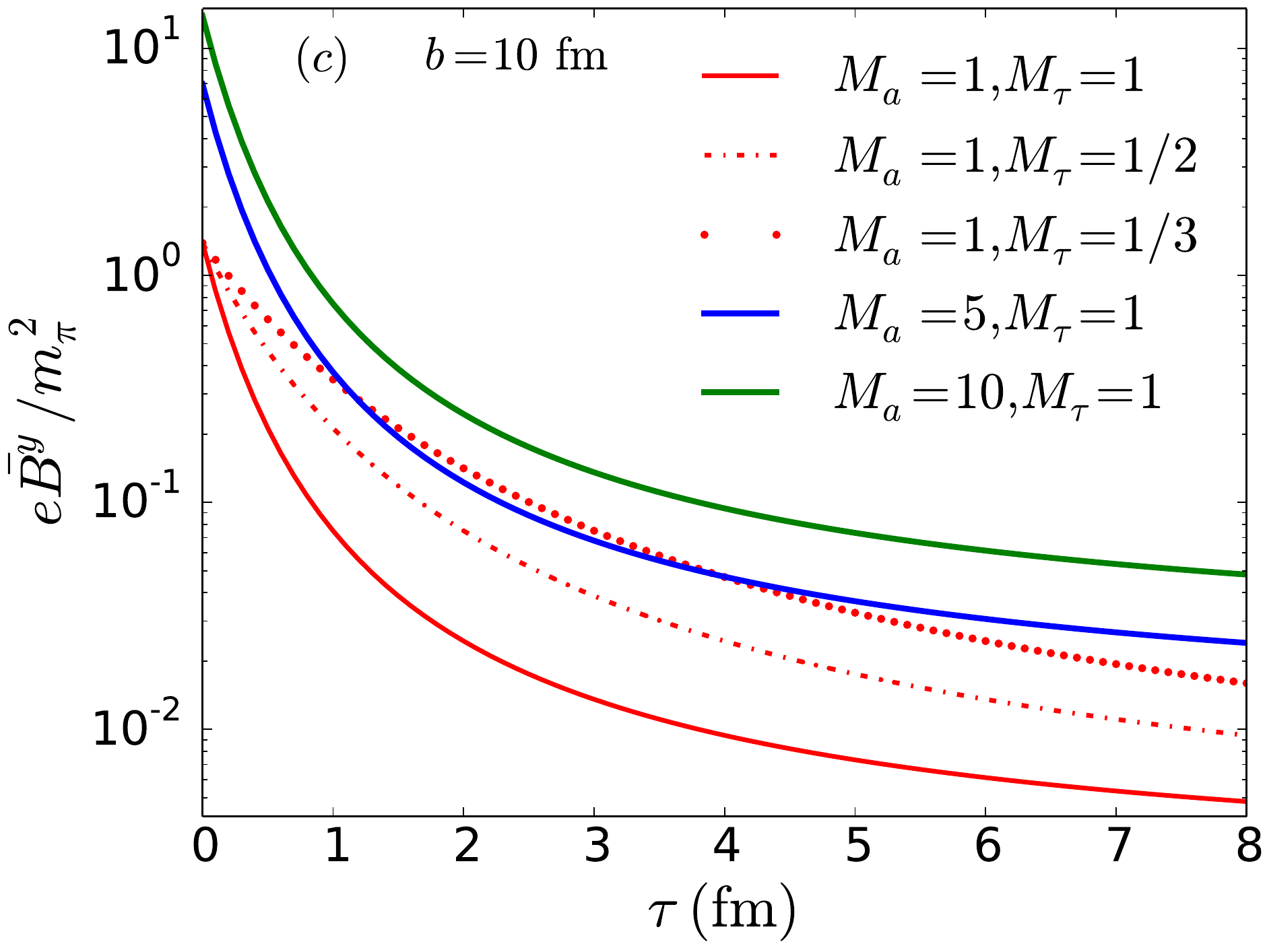}
\caption{Evolution of $e\bar{B}^y$, normalised to the pion mass squared at $x =
  y = 0$. $(a)$ Evolution of $e\bar{B}^y$ in vacuum, for $b = 0\,{\rm fm}$
  collisions (red circles and line), and for $b = 10$ fm collisions (blue
  triangles and line). The values corresponding to symbols are taken from
  Ref.~\cite{Deng:2012pc}, the lines are fits, respectively. $(b)$
  Evolution of $e\bar{B}^y$ in medium with a finite conductivity for $b =
  10\,{\rm fm}$ collisions, red open circles are from
  Ref.~\cite{Tuchin:2013apa}, the black solid line is a fit. $(c)$ The
  same as middle panel, but for various values of the fit parameters.}
\label{fig:Time_evo_B}
\end{figure*}
 
At each time-step the conserved quantities $T^{\tau\tau}, T^{x\tau},$ and
$T^{y\tau}$ are evolved to the next time-step using the SHASTA
algorithm. In order to find the primitive variables
$\varepsilon,p, v^{x},v^{y}$ from the time-evolved conserved quantities we use
{the} following algorithm \cite{Rischke:1995ir}. First we define the quantities
\begin{align}
\mathcal{E}  := & T^{\tau\tau} =  w \gamma^{2}-p_B- B^2 (b^{\tau})^{2}\;,  \\
\mathcal{M}^x := & T^{\tau x}  = w \gamma^{2}v^{x}- B^2 b^{\tau}b^{x}\; , \\
\mathcal{M}^y := & T^{\tau y}  = w \gamma^{2}v^{y} -B^2 b^{\tau}b^{y}\; .
\end{align}
Note that the momentum flow vector $\boldsymbol{\mathcal{M}} =
\left(\mathcal{M}^x,\mathcal{M}^y\right)$ is not always parallel to the fluid velocity
vector $\boldsymbol{v} = \left(v^x,v^y\right)$ and thus we cannot apply
the algorithm given in the original ``\texttt{AZHYDRO}'' code to find the
new velocity. To counter this problem we next introduce the new
quantities
\begin{eqnarray}
\mathcal{E}^{'} := & \mathcal{E}+ B^2 (b^{\tau})^2 = w \gamma^2 -p_B\;, \\
{\mathcal{M}^{x}}^{'} := & \mathcal{M}^{x} + B^2 b^{\tau}b^x  =  w \gamma^2v^{x}\;, \\
{\mathcal{M}^{y}}^{'} := & \mathcal{M}^{y} + B^2 b^{\tau}b^y  =  w \gamma^2v^{y}\;,
\end{eqnarray}
where the new three-vector $\boldsymbol{\mathcal{M}^{'}} = ({\mathcal{M}^x}^{'},{\mathcal{M}^y}^{'})$ is
always parallel to $\boldsymbol{v}$. As a result, we can now apply the
well known technique (given below) of finding primitive variables at each
time-step. More specifically, after defining $\mathcal{M}^{'}:=|\boldsymbol{\mathcal{M}}^{'}|$ and
$v:=|\boldsymbol{v}|$, we can write
\begin{align}
\mathcal{M}^{'} = & \left(\mathcal{E}^{'}+p_B \right)v\; , \\ 
\varepsilon  = & \; \mathcal{E}^{'}-\mathcal{M}^{'}v -\frac{B^{2}}{2}\; ,
\label{eq:varEps}
\end{align}
and use the above expressions to replace $\varepsilon$ in
$p(\varepsilon)$ to finally obtain
\begin{equation}
\label{eq:new_v}
v = \left. \frac{\mathcal{M}^{'}}{\mathcal{E}^{'}+p(\varepsilon)} \right|_{\varepsilon =
  \mathcal{E}^{'}-\mathcal{M}^{'}v-{B^{2}}/{2}} \; .
\end{equation}
For given values of $\mathcal{E}^{'},\mathcal{M}^{'}$, and $B^{2}$, Eq. \eqref{eq:new_v} can
be solved iteratively for the velocity $v$, which, once known, allows us
to compute $\varepsilon$ from Eq.~(\ref{eq:varEps}). Finally, the
distinct components $v^x$ and $v^y$ can be obtained from the collinearity
of $\boldsymbol{\mathcal{M}^{'}}$ and $\boldsymbol{v}$.

\subsection{Initial data}
\label{sec:ID}

Obviously, in order to solve the system of coupled partial differential
equations (\ref{eq:EMConsT})--(\ref{eq:EMConsY}) a set of initial
conditions needs to be specified. In particular, at the initial time of
the hydrodynamical evolution, which we {choose as} $\tau_0 = 0.6\,{\rm
  fm}$, we set $v^x=v^y=0$, while the initial energy density in the
transverse plane is obtained from the Glauber model via the following
two-component form
\begin{equation}
\varepsilon\left(x,y,b\right) =
\varepsilon_0\left[x_{h}N_{part}\left(x,y,b\right)+
  (1-x_{h})N_{coll}\left(x,y,b\right)\right]\;.
\end{equation}
Here, $N_{part}\left(x,y,b\right)$ and $N_{coll}\left(x,y,b\right)$ are
the transverse profiles of the average number of participants and the
average number of binary collisions, respectively, both calculated within
a Glauber model for a given impact parameter $b$. The fraction of hard
scattering $x_h$ is important to explain {the} centrality dependence of the
average charged hadron multiplicity. Since we will not compare our result
to experimental data, we take $x_h = 0.25$ in all cases
considered.

\subsection{Magnetic-field evolution}
\label{sec:MFE}

\begin{figure*}
\includegraphics[width=0.45\textwidth]{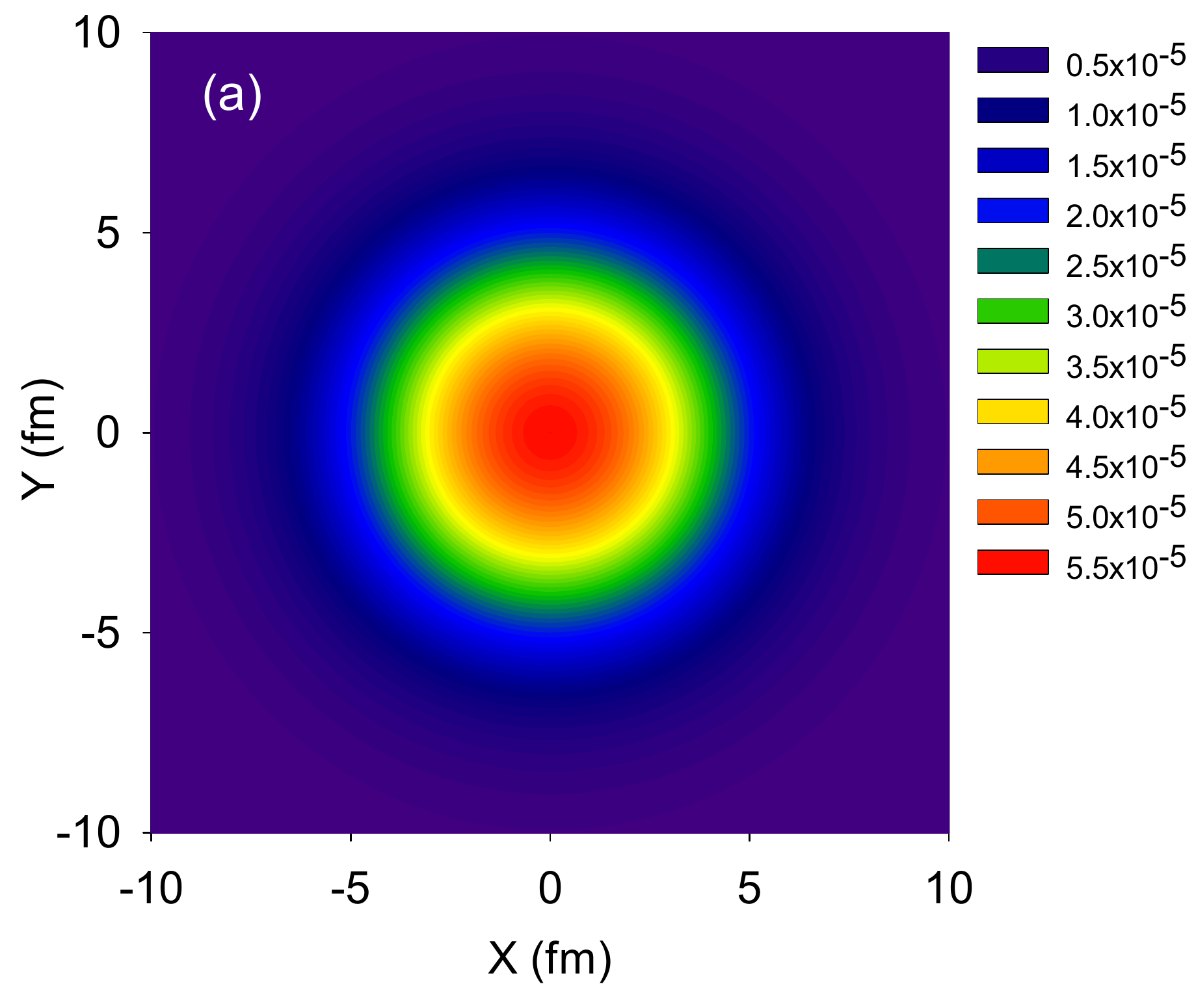}
\includegraphics[width=0.45\textwidth]{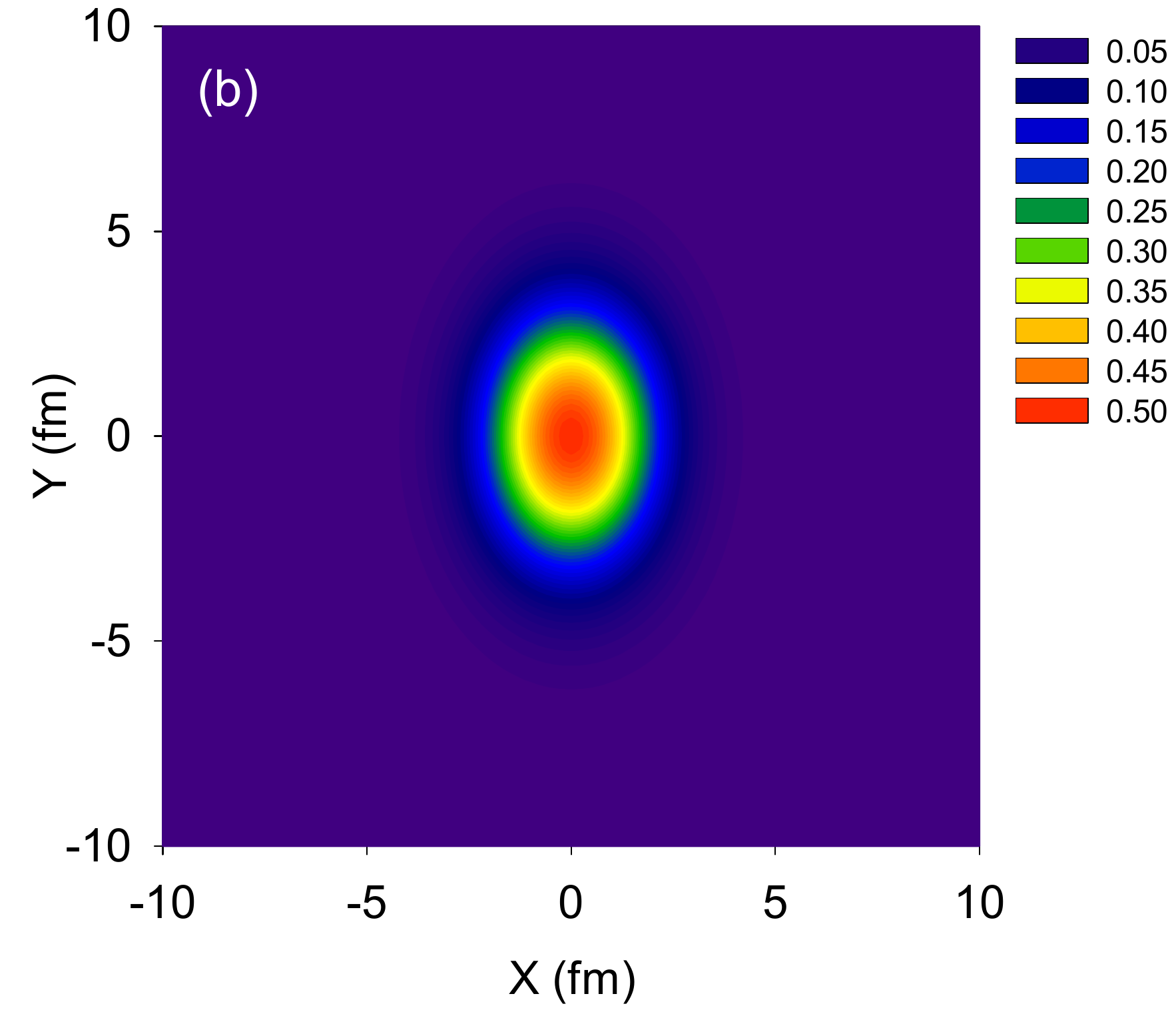}
\caption{Magnetic-field energy density in the transverse plane at $\tau =
  0\,{\rm fm}$. The panel $(a)$ refers to collisions with $b = 0\,{\rm
    fm}$, while panel $(b)$ to collisions with $b = 10\,{\rm fm}$.}
\label{fig:By2Gaussian}
\end{figure*}

In a fully consistent solution of the MHD equations with appropriate
boundary conditions the induction equation would provide the evolution of
the magnetic {field} as a result of the dynamics of the magnetized
flow. However, as mentioned in Sec.~\ref{sec:introduction}, we here
employ a reduced set of MHD equations, and the evolution of the external
magnetic field is taken to follow some suitably defined function in space
and time. Inspired by a previous study \cite{Roy:2015coa}, we use the
following parametrized form in space and time for the $y$-component of
the magnetic field
\begin{eqnarray}
\frac{e\bar{B}^{y}\left(x,y,\tau\right)}{m_\pi^2} =  
f\left(\tau\right) 
\exp\!\left[-\frac{\left(x-x_0\right)^2}{4\sigma^2_x} -
\frac{\left(y-y_0\right)^2}{4\sigma^2_y}\right]\;, \nonumber \\
\label{eq:ByGaussian}
\end{eqnarray}
In all cases considered we center the Gaussian in
{Eq.~\eqref{eq:ByGaussian}} at $x_0 = y_0 = 0$ and use $\sigma_x,\,
\sigma_y$ to set the widths of the Gaussian in $x$- and $y$-direction,
respectively. For an impact parameter $b = 0$, we use $\sigma_x =
\sigma_y= 3.5\,{\rm fm} $, while for $b = 10\,{\rm fm}$, we set $\sigma_x
= 1.5\,{\rm fm}$ and $\sigma_y = 2.2\,{\rm fm}$. The corresponding
magnetic energy densities at $\tau=0$ are shown in
Fig.~\ref{fig:By2Gaussian} for the cases of $b=0$ (left panel) and
$b=10\,{\rm fm}$ (right panel).

The evolution of the magnetic field in the QGP is not well known. In
vacuum, the decay time of the magnetic field is inversely proportional to
the $\sqrt{s_{\rm NN}}$ of the collision \cite{Deng:2012pc}. However,
several studies have shown that the QGP possesses a nonzero
temperature-dependent electrical conductivity
~\cite{Gupta:2003zh,Greif:2014oia,Finazzo:2013efa}. In this case, the
decay of the magnetic field can be substantially
delayed~\cite{Tuchin:2011jw,Li:2016tel}.

In view of these considerations and uncertainties, we here employ a
function of proper time only, i.e., $f\left(\tau\right)$ in
Eq.~\eqref{eq:ByGaussian}, as a fully phenomenological ansatz for a
reasonable parametrization of the evolution of the magnetic field $\bar{B}^y$,
distinguishing the case in which the field is in vacuum from when it is
in a QGP.

\begin{enumerate}
\item[(i)] In \emph{vacuum} we parametrize the evolution of the
  magnetic field as in Ref.~\cite{Deng:2012pc}, so that for $b =
  0\,{\rm fm}$ collisions
\begin{equation} 
\label{eq:param1}
f\left(\tau\right) = \frac{1}{a_1+b_1 \tau}\; , 
\end{equation}
and for $b = 10\,{\rm fm}$ collisions 
\begin{equation} 
\label{eq:param2}
f\left(\tau\right) = a_2 e^{b_2/(\tau + c_2)}\;.
\end{equation}

Adjusting the constants in these parametrizations to the data given in
Ref.~\cite{Deng:2012pc}, we obtain $a_1 = 78.2658$, $b_1 = 79.5457\;
\rm{fm^{-1}}$, $a_2 = 1.357\times 10^{-4} $, $b_2 = 3.1031\; \rm{fm}$,
and $c_2 = 0.2483\; \rm{fm}$. The data {are} shown by the symbols in
Fig.~\ref{fig:Time_evo_B} (a), while our parametrizations
(\ref{eq:param1}) and (\ref{eq:param2}) are given by the lines in that
figure. From now on we denote these parametrizations as {\it
  ``Parvac''}, since they are valid in vacuum.

\item[(ii)] In a QGP with nonzero electrical conductivity we
  parametrize the evolution of the magnetic field as in
  Ref.~\cite{Tuchin:2013apa} [see Fig.\ 3 of Ref.~\cite{Tuchin:2013apa}]
\begin{equation} 
f\left(\tau\right) = M_a a_3 e^{b_3/(M_\tau \tau + c_3)}\;.
\label{eq:Parmed}
\end{equation}
We denote this parametrization as {\it ``Parmed''}. Data from
Ref.~\cite{Tuchin:2013apa} are shown in Fig.~\ref{fig:Time_evo_B} (b).
We fit these data setting $M_a = M_\tau = 1$ and adjusting the constants,
giving $a_3 = 1.99\times 10^{-3}$, $b_3 = 8.1306\; \rm{fm}$, and $c_3 =
1.2420\; \rm{fm}$. We note that at late times, \ie for $\tau\geq 5\,{\rm
  fm}$, the fit (black line) overestimates the corresponding data points
(open red circles), but also that the magnetic field at this time is
already two orders of magnitude smaller than its initial value, so that
this mismatch is likely not dynamically important.

As an extension of the space of parameters we have also studied variations
of the parametrization (\ref{eq:Parmed}) by changing the constants $M_a$
and $M_\tau$. Since varying $M_a$ changes the value of $\bar{B}^y$ at $\tau =
0$, we have considered $M_a = 1, 5$, and $10$, which corresponds to
$e\bar{B}^y/m_{\pi}^2 \sim 1, 5$, and $10$ at $\tau = 0$,
respectively. Furthermore, the decay rate has been varied by using
different values of $M_\tau$ and for each value of $M_a$ we use three
different values, namely, $M_\tau = 1, 1/2$, and $1/3$.

\end{enumerate}


\section{Results}
\label{sec:results}

In order to measure the effect of a strong magnetic field we investigate
the evolution of the ``momentum anisotropy'' of the fluid flow in Au+Au
collisions and defined as
\begin{equation}
\varepsilon_p(\tau) := \frac{\langle T^{xx} - T^{yy}\rangle}{\langle
  T^{xx} + T^{yy}\rangle}\;,
\end{equation}
where $\langle \cdots \rangle$ denotes the energy-density
weighted average over the transverse plane at proper time $\tau$, \ie
for a generic component
\begin{equation}
\langle T^{ij} (\tau) \rangle := \frac{\int dx\, dy\, \varepsilon(x,y,\tau)\,
  T^{ij}(x,y,\tau)} {\int dx \, dy\, \varepsilon(x,y,\tau)}\;.
\end{equation}

\begin{figure}
\includegraphics[width=0.8\columnwidth]{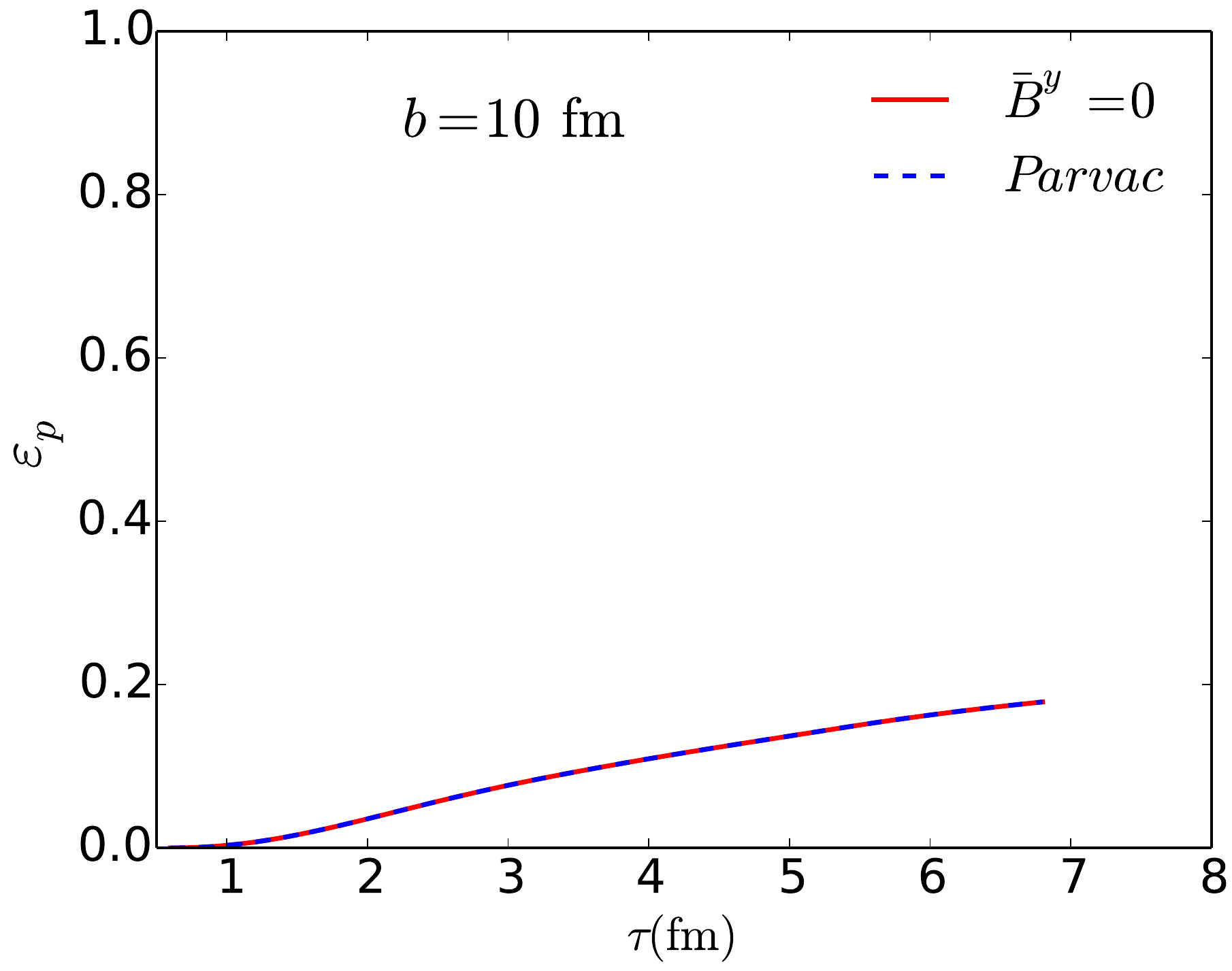}
\caption{{Evolution of the momentum anisotropy $\varepsilon_p$ for $b =
    10\,{\rm fm}$ collisions when the magnetic field is taken to be zero
    (solid red line) or to follow the {\it Parvac\/} parameterisation
    (dashed blue line).}}
\label{fig:Meccen_XG}
\end{figure}

The momentum anisotropy is a particularly interesting quantity to study
since an azimuthally asymmetric energy-density distribution in the
transverse plane in non-central collisions is expected to give rise to
stronger pressure gradients along the $x$-direction than along the
$y$-direction, at least in our geometrical setup. In turn, since pressure
gradients drive the fluid flow, a momentum anisotropy of this type is
directly related to a higher flow velocity along the $x$-direction than
along the $y$-direction. In Ref.~\cite{Kolb:1999it} it was shown that
$\varepsilon_p$ at freeze-out is directly related to the
transverse-momentum squared $\left(p^{2}_{T}\right)$ {weighted} elliptic
flow of pions. Thus, any change in $\varepsilon_p$ also indicates a
possible change in the elliptic flow of hadrons and the following results
corroborate this expectation.

\begin{figure*}
\includegraphics[width=0.33\textwidth]{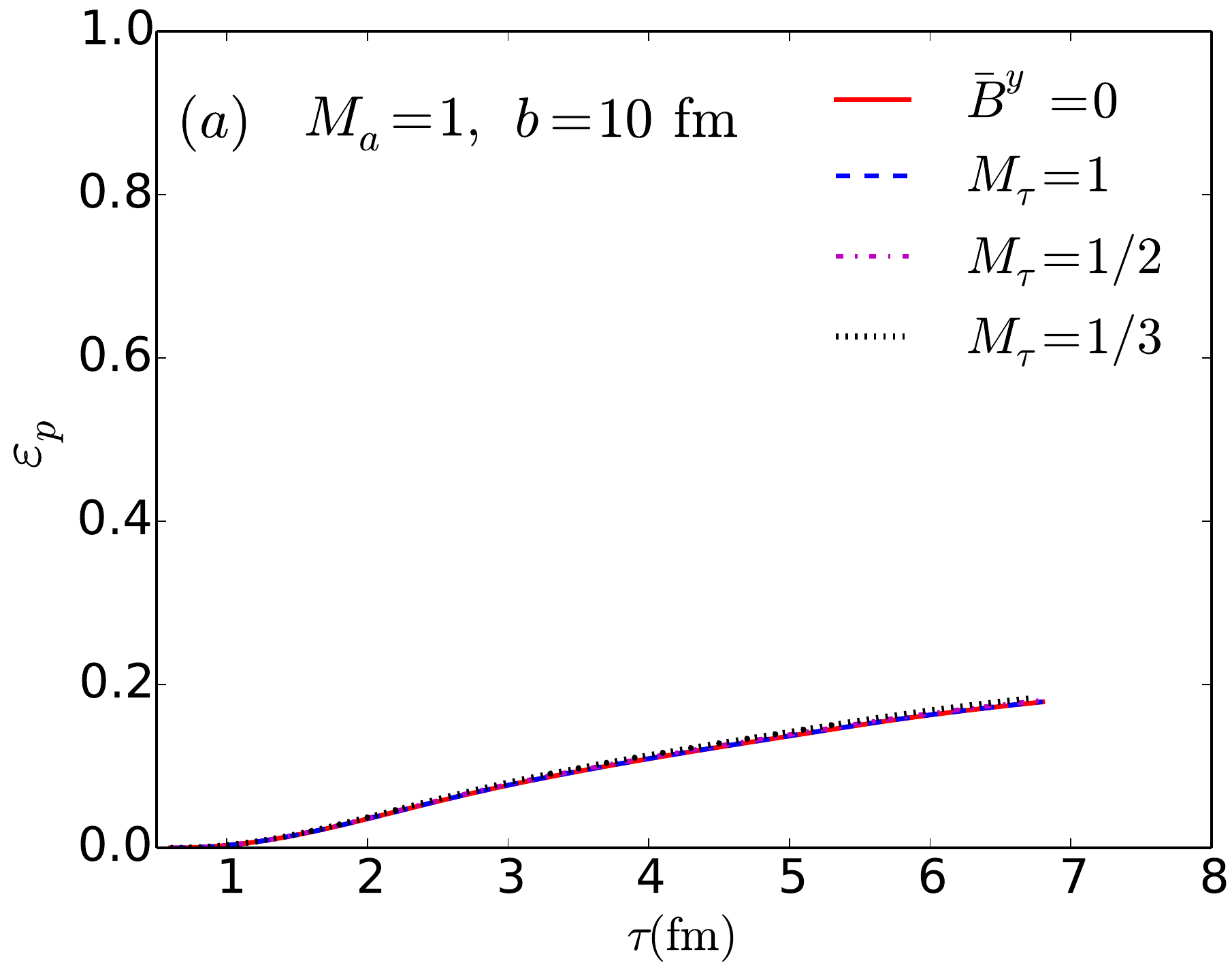}
\includegraphics[width=0.33\textwidth]{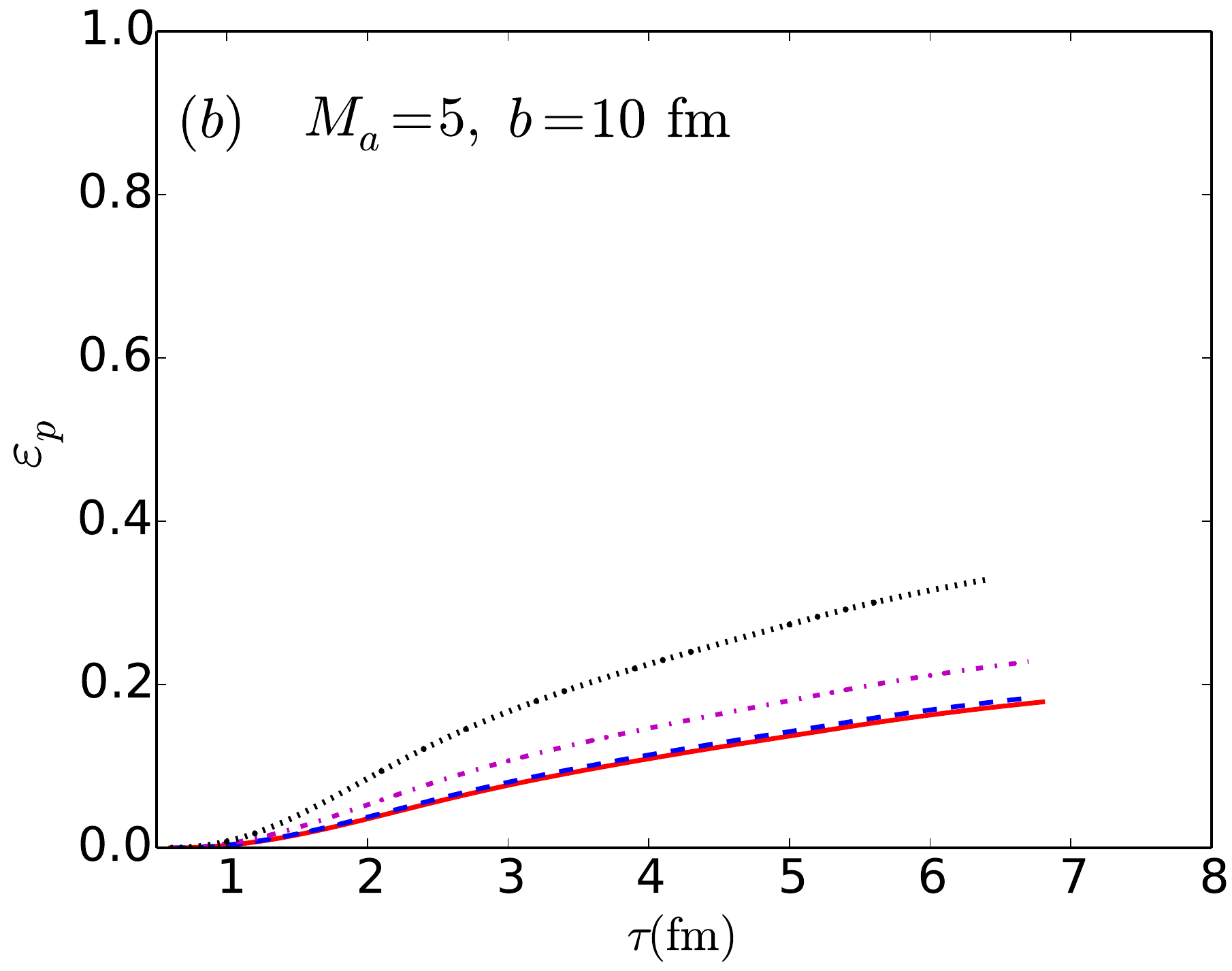}
\includegraphics[width=0.33\textwidth]{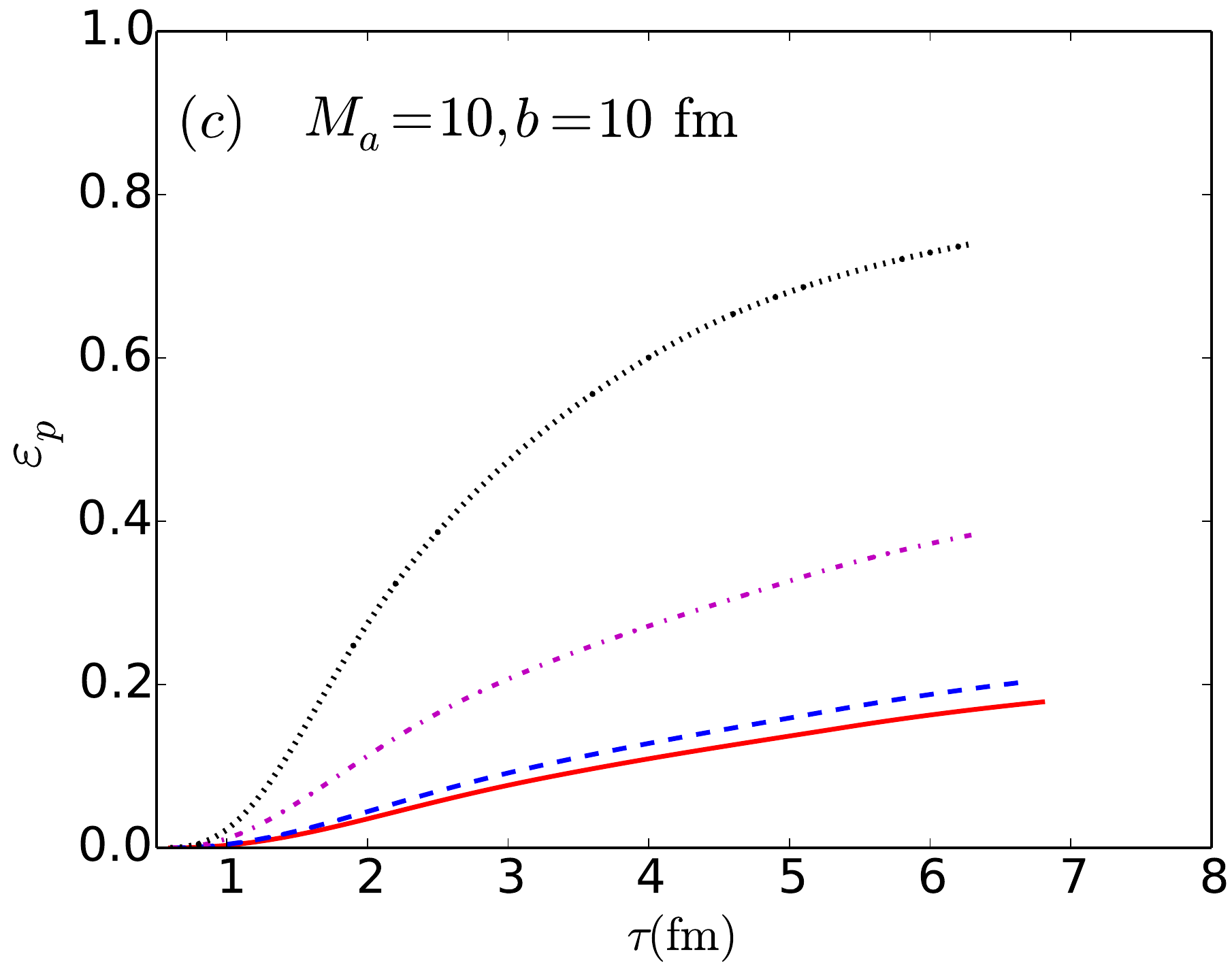}
\caption{Evolution of the momentum anisotropy for $b = 10\,{\rm fm}$
  collisions for the {\it Parmed\/} case. $(a)$ The solid red line
  corresponds to the result for zero magnetic field, the dashed blue,
  dash-dotted magenta, and dotted black lines correspond to results with
  external magnetic field for $M_\tau = 1, 1/2$ , and $1/3$,
  respectively; in all cases $M_a = 1$. $(b)$ The same as in (a), but for
  $M_a = 5$. $(c)$ The same as in $(a)$, but for $M_a = 10$.}
\label{fig:b10MomSpatEccen}
\end{figure*}

As an initial test of the numerical infrastructure we have considered the
simplified but also physically less interesting case of central
collisions, \ie $b=0$. In this case, the symmetry of the system yields
$\varepsilon_{p} = 0$ at all times in a purely hydrodynamical
flow. Actually, this result applies also in the presence of a magnetic
field, since the magnetic-field contribution in the $x$-direction is
expected to be the same as the one in the $y$-direction, at least when
$b=0$. However, our numerical setup, in which only $\bar{B}^y$ is switched on,
does not allow us to validate this behaviour, but we have verified that
the growth of $\varepsilon_{p} $ is nevertheless extremely small, being
$\varepsilon_{p} \lesssim 10^{-6}$ for a Parvac parametrization and
$\varepsilon_{p} \lesssim 2 \times 10^{-3}$ for a Parmed parametrization
with $M_a=5$.

On the other hand, for peripheral collisions one expects an anisotropy to
develop already from the underlying asymmetric hydrodynamical flow. This
anisotropy can then be further amplified if a magnetic field is present.
Figure~\ref{fig:Meccen_XG} shows the growth of such anisotropy by
reporting the evolution of $\varepsilon_p$ for a collision with
$b=10\,{\rm fm}$. Shown with a solid red line is the purely
hydrodynamical evolution (\ie with zero magnetic field), while the dashed
blue line refers to the Parvac parametrization. Clearly the two curves
are very similar and this is essentially because with the parametrization
\eqref{eq:param2} the magnetic field {is} effectively very small,
$e\bar{B}^y/m_\pi^2 \lesssim 10^{-2}$ [\cf Fig.~\ref{fig:Time_evo_B} (a)].

The evolution of the momentum anisotropy $\varepsilon_p$ for the case of
collisions with $b = 10\,{\rm fm}$ and when the magnetic field is evolved
using the {\it Parmed\/} parametrization is shown in
Fig.~\ref{fig:b10MomSpatEccen}. More specifically,
Fig.~\ref{fig:b10MomSpatEccen} (a) corresponds to case {where} the
initial magnetic-field amplitude is $M_a = 1$, \ie when the magnetic
field at $\tau = 0$ is set to be $e\bar{B}^y \sim m_{\pi}^2$. The solid red
line corresponds to the case without magnetic field, while the dashed
blue, dash-dotted magenta, and the dotted black lines correspond to
$M_\tau = 1, 1/2$, and ${1/3}$, respectively. The evolution is shown up
to freeze out, that is when {the temperature is nowhere larger than
  $T_f=130\,{\rm MeV}$.}

A rapid inspection of {Fig.~\ref{fig:b10MomSpatEccen} (a)} reveals that a
visible change in $\varepsilon_p$ is seen only when the magnetic field
decays very slowly, \ie for $M_{\tau}=1/3$ (dotted black line). Under
these conditions one is induced to conclude that the influence of the
magnetic field is very limited and that the momentum anisotropy remains
small, with a relative variation relative to the purely hydrodynamical
case {of} $|1 - \varepsilon_p/\varepsilon_p(\bar{B}^y=0)| \lesssim 3\times
10^{-2}$. However, because the common expectation is that the initial
magnetic field in $b = 10\,{\rm fm}$ Au+Au collisions can be
substantially larger than $m_\pi^2$, {Fig.~\ref{fig:b10MomSpatEccen} (b)}
reports the evolution of the momentum anisotropy for a larger initial
magnetic field, \ie $M_a = 5$ or $e\bar{B}^y\simeq 5 \,m_{\pi}^2$ at $\tau =
0$. In this case, in fact, even for the most rapid decay of the magnetic
field, \ie $M_{\tau} = 1$, the momentum anisotropy $\varepsilon_p$ is
larger when compared to the case of zero magnetic field; the largest
relative difference in this case is $|1 -
\varepsilon_p/\varepsilon_p(\bar{B}^y=0)| \sim 0.8$ and is obviously obtained
for $M_{\tau}=1/3$. Finally, as can be seen from
{Fig.~\ref{fig:b10MomSpatEccen} (c)}, a much higher initial value of the
magnetic field (\ie $M_a = 10$) increases $\varepsilon_p$ even more, with
a relative difference that can now be $|1 -
\varepsilon_p/\varepsilon_p(\bar{B}^y=0)| \sim 3.2$ for $M_{\tau}=1/3$.

As mentioned earlier, the elliptic-flow coefficient $v_2$ of charged
hadrons is directly proportional to the momentum anisotropy
$\varepsilon_p$, so that we expect also a noticeable change of $v_2$ due
to the magnetic field. For demonstration purposes, we show here $v_2$ of
$\pi^{-}$ only\footnote{Note that elliptic-flow coefficient $v_2$ for
  $\pi^+$ would be identical, since any effect of the magnetic field
  after freeze-out is neglected.}. Since we are not trying to match
experimental data, the input parameters for simulations are not adjusted
to reproduce any experimentally measured charged-hadron
multiplicity. However, we do use realistic values for the input
parameters corresponding to Au+Au collisions at $\sqrt{s_{\rm NN}} =
200\,{\rm GeV}$. More specifically, at the initial time $\tau_0 =
0.5\,{\rm fm}$ and for central collisions ($b = 0\,{\rm fm}$) we set the
central energy density to be $\varepsilon =50\,{\rm GeV\ fm}^{-3}$ and
consider a constant freeze-out temperature of $120\,{\rm MeV}$. However,
neither resonance decays nor viscous corrections are taken into account.

\begin{figure*}
\includegraphics[width=0.33\textwidth]{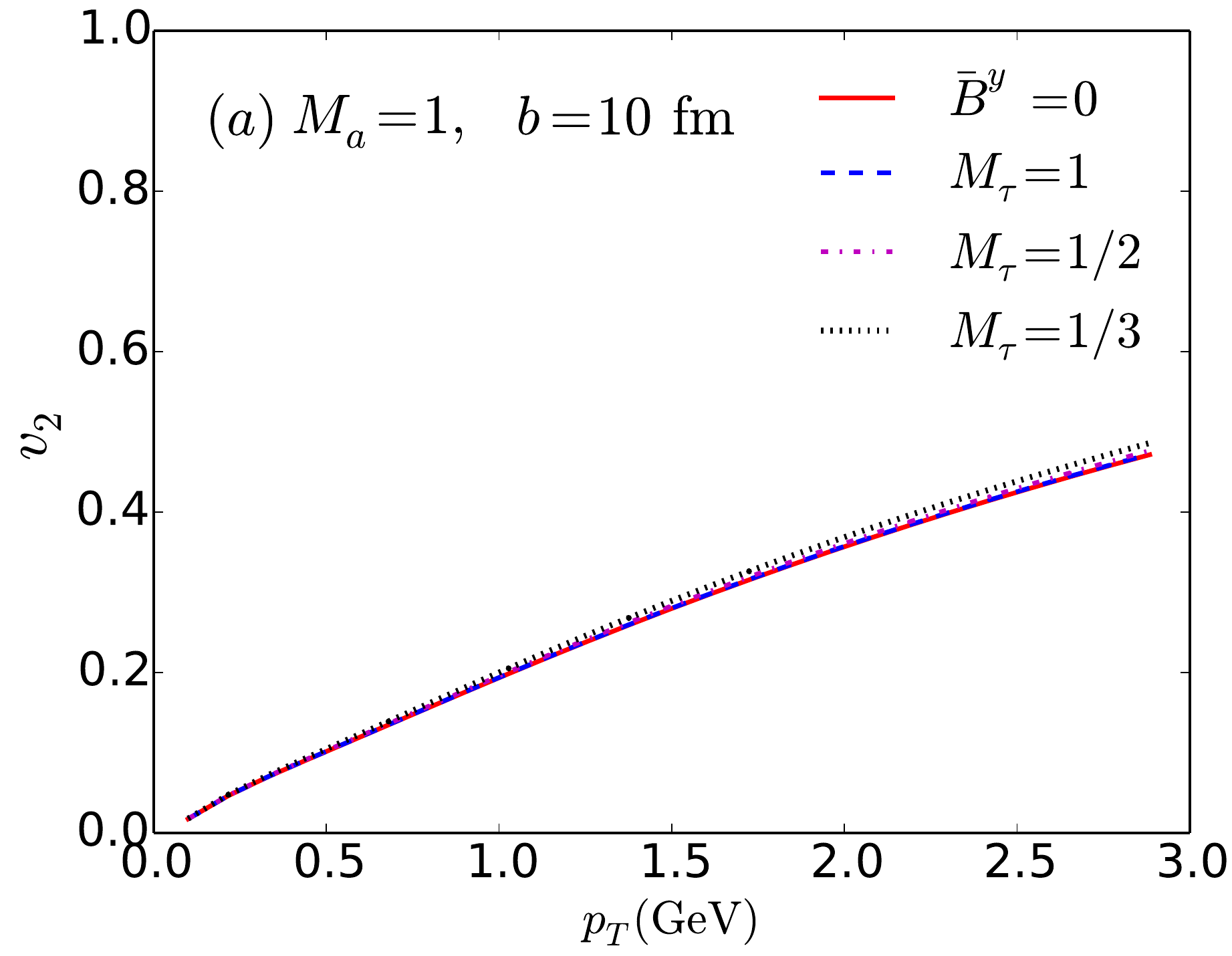}
\includegraphics[width=0.33\textwidth]{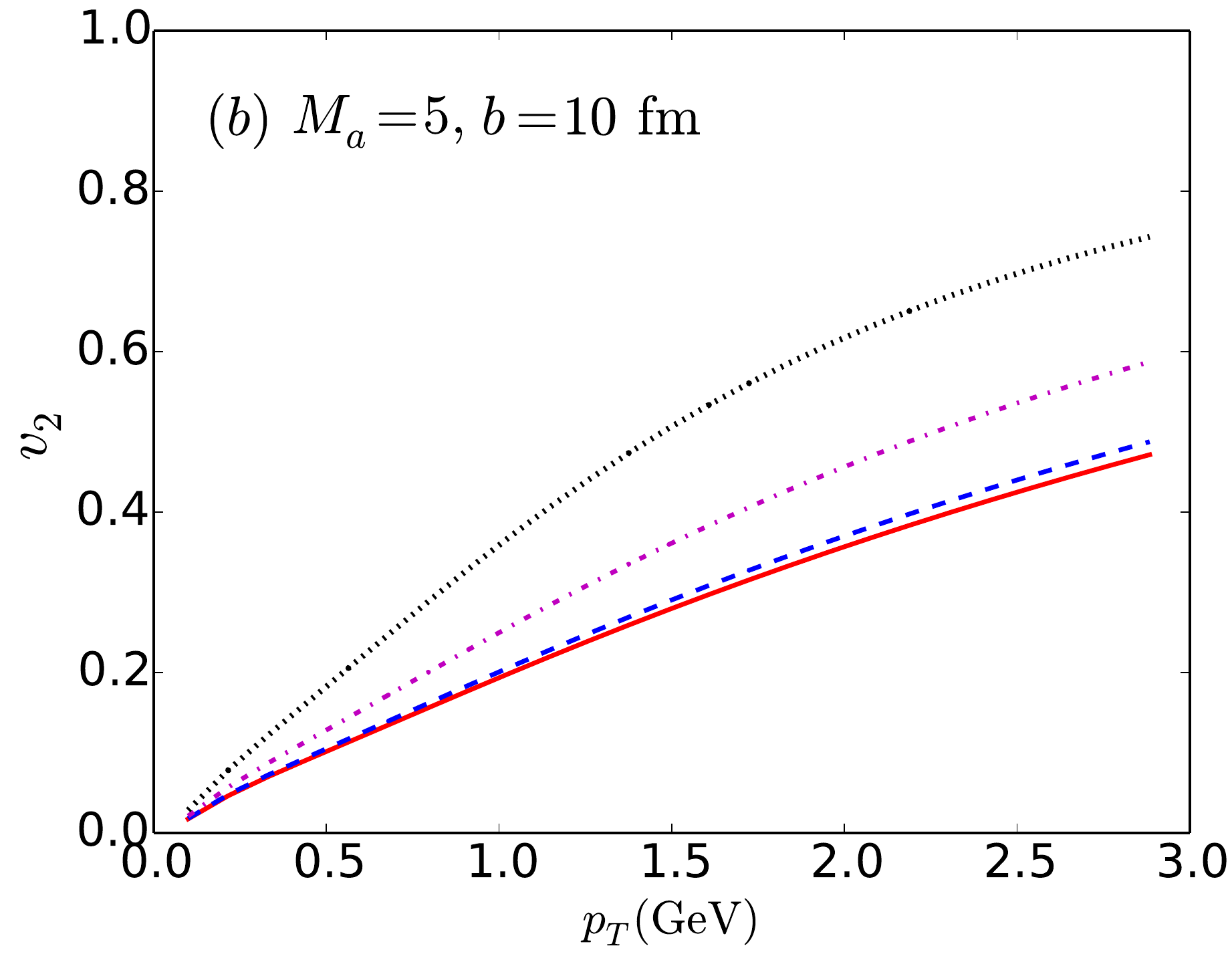}
\includegraphics[width=0.33\textwidth]{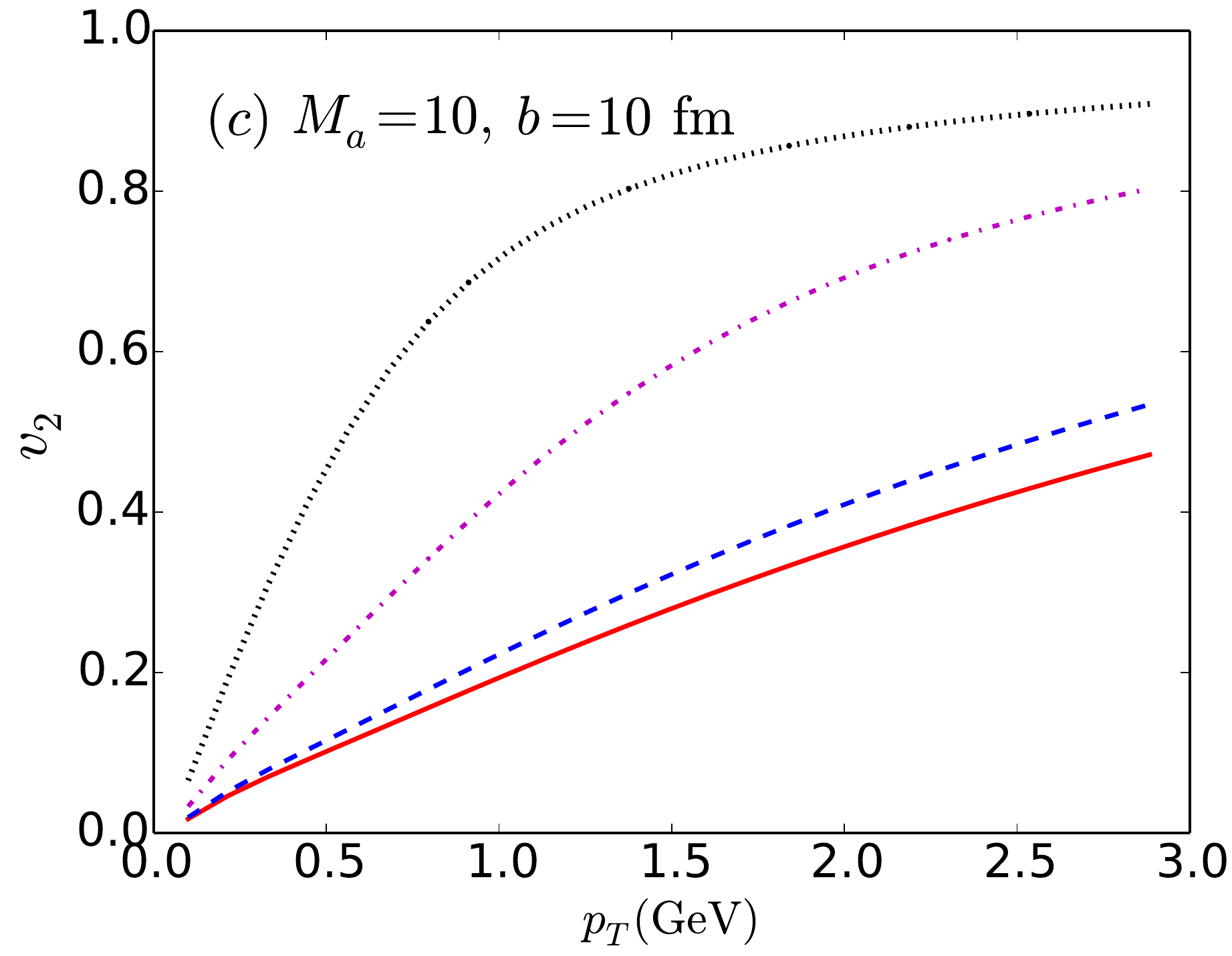}
\caption{The elliptic-flow coefficient $v_2$ for $\pi^{-}$ as a function
  of transverse momentum $p_T$ for $b = 10\,{\rm fm}$ collisions. $(a)$
  The solid red line corresponds to the result for zero magnetic field,
  the dashed blue, dash-dotted magenta, and dotted black lines correspond
  to results for an external magnetic field with $M_\tau = 1, 1/2$, and
  $1/3$, respectively. All results are obtained for $M_a = 1$. $(b)$ The
  same as in $(a)$, but for $M_a = 5$. $(c)$ The same as in $(a)$, but
  for $M_a = 10$.}
\label{fig:b10v2}
\end{figure*}

Figure~\ref{fig:b10v2} shows the elliptic-flow coefficient $v_2$ of
$\pi^{-}$ as a function of the transverse momentum $p_T$ for non-central
collisions with $b = 10\,{\rm fm}$. Different lines refer to the same set
of conditions as in Fig.~\ref{fig:b10MomSpatEccen}, namely, the solid red
line corresponds to the result for zero magnetic field, the dashed blue,
dash-dotted magenta, and dotted black lines correspond to results with
external magnetic field for $M_\tau = 1, 1/2$ , and $1/3$,
respectively. In analogy with what discussed for the {momentum
  anisotropy}, it is clear from Fig.~\ref{fig:b10v2} that changes in
$v_2$ are noticeable only when either the initial magnetic field is large
or when the magnetic field decay is substantially delayed. For the
largest initial value of the magnetic field considered here, \ie for $e
\bar{B}^y \simeq 10 \,m_{\pi}^2$, we notice a considerable enhancement of
the elliptic-flow coefficient, which can become as large as $v_2 \lesssim
0.9$ for $p_T \sim 2.5\,{\rm GeV}$ [\cf dotted black line in
{Fig.~\ref{fig:b10v2} (c)}]. A smaller initial magnetic field, \ie $e
\bar{B}^y \simeq 5 \,m_{\pi}^2$, leads to a smaller increase of the
elliptic-flow coefficient, which however remains rather large, with $v_2
\lesssim 0.7$ for $p_T \sim 2.5\,{\rm GeV}$ [\cf dotted black line in
Fig.~\ref{fig:b10v2} (b)], thus highlighting that quite realistic values
of the magnetic field can have a considerable impact on the ellipticity
of the flow of particles. Overall, these results and their implications
for the understanding of the physics of ultrarelativistic heavy-ion
collisions clearly call for the extension of this study towards a fully
self-consistent MHD treatment of the evolution of hot and dense strongly
interacting matter created in heavy-ion collisions, following the spirit
of the work in Ref.~\cite{Inghirami:2016iru}.

\section{Conclusions}
\label{sec:summary}

We have investigated the effect of a strong external magnetic field on
the evolution of matter created in $\sqrt{s_{\rm NN}} = 200\,{\rm GeV}$
Au+Au collisions within a $2+1$ dimensional reduced-MHD description. In
particular, we have assumed that the external magnetic field has only a
non-vanishing component transverse to the reaction plane (\ie it is
aligned with the $y$-direction) and we have employed the spacetime
variation suggested in Refs.~\cite{Deng:2012pc,Tuchin:2013apa}. Overall
and on average, we found no visible changes in the fluid-velocity
profile when comparing the magnetic-field decays in vacuum with the case
in which the magnetic field is actually zero.

On the other hand, a substantial change in the fluid velocity and,
consequently, in the elliptic-flow coefficient $v_2$ of $\pi^{-}$ is
observed when {the magnetic field is sufficiently large, \ie for $e\bar{B}^y
  \gtrsim 5\, m_\pi^2$, or when a nonzero electrical conductivity of the
  QGP is accounted for such that} it decays slowly, \ie for $M_{\tau}
\gtrsim 1/2$. Under these conditions, the momentum anisotropy shows a
relative variation relative to the purely hydrodynamical case {of} $|1 -
\varepsilon_p/\varepsilon_p(\bar{B}^y=0)| \gtrsim 1/2$, while the elliptic-flow
coefficient can become as large as $v_2 {\sim} 0.7$ for $p_T \sim
2.5\,{\rm GeV}$ (all of the values reported refer to an initial magnetic
field strength $e\bar{B}^y \simeq 5\, m_\pi^2$). 

Our results are obtained under some simplifying assumptions: (1) We have
used an analytic prescription for the magnetic-field evolution, but the
latter should really be the result of a self-consistent solution of the
full set of ideal-MHD equations~\cite{Inghirami:2016iru}. (2) We have
considered event-averaged values for the initial energy density and the
magnetic field, but both of them fluctuate event to event in
reality. Indeed, a previous study \cite{Roy:2015coa} has shown that
because of the event-by-event fluctuations of both the magnetic energy
density and of the fluid energy density, in some cases the ratio of these
two quantities can be $\sim\,1$. In such cases, the magnetic field will
have a larger effect than considered here. (3) We have neglected the $x$
component of the magnetic field as we expect that $\bar{B}^x \ll
\bar{B}^y$ in the present geometrical setup. Although this is a good
approximation for peripheral collisions, in central collisions
$\bar{B}^x$ is of the same order as $\bar{B}^y$ and one needs to consider
both. (4) We have considered a decay of the magnetic field pertaining to a
constant electrical conductivity \cite{Tuchin:2013apa}. However, one
should use the appropriate temperature-dependent electrical conductivity
of the QGP. (5) We have considered the case of vanishing magnetization, but, depending
on the magnetic properties of the QGP and the hadronic phase, a nonzero
magnetization of the medium needs to be accounted for in the full
energy-momentum tensor, as some recent preliminary studies show that this
could also affect the QGP evolution \cite{Pu:2016ayh, Pang:2016yuh}. (6)
We have considered here only perfect fluids \cite{Rezzolla_book:2013},
but it is important to take into account also dissipative corrections to
the fluid evolution. {Nonzero magnetic fields will} have an impact on the
{value of the shear viscosity-to-entropy density ratio} $\eta_{\rm sh}/s$
extracted from a comparison to experimental data, as was also speculated
in some previous studies \cite{Tuchin:2011jw, Mohapatra:2011ku}.

Overall, we regard the present study as of exploratory nature. In
addition to the considerations made above and together with a systematic
exploration of the input parameters, our work will need to be extended in
a number of ways. These include: the study of the corrections to the
final particle spectra due to the magnetic field at and after freeze-out,
the study of several other experimental observables, \eg charge-dependent
azimuthal correlations and soft-photon production
\cite{Bloczynski:2012en, Bloczynski:2013mca, Deng:2014uja,Deng:2016knn}, as well as
the investigation of smaller collision energies, where the decay of the
magnetic field is slower and thus its impact on the fluid evolution is
expected to be more pronounced.

\section*{Acknowledgements}
\noindent It is a pleasure to thank Gergely Endr\"odi, Xu-Guang Huang, Igor Mishustin,
          Long-Gang Pang, and Hannah Petersen for discussions and
          comments. VR is supported by the DST-INSPIRE faculty research
          grant. SP is supported by a JSPS post-doctoral fellowship for
          foreign researchers. DHR is partially supported by the High-end
          Foreign Experts project GDW20167100136 of the State
          Administration of Foreign Experts Affairs of China. Support
          also comes from ``NewCompStar'', COST Action MP1304, and from
          the LOEWE program HIC for FAIR.


\begin{thebibliography}{1}
 

\bibitem{Deng:2012pc} 
  W.~T.~Deng and X.~G.~Huang,
  ``Event-by-event generation of electromagnetic fields in heavy-ion collisions,''
  Phys.\ Rev.\ C {\bf 85}, 044907 (2012).
  
\bibitem{Tuchin:2013apa} 
  K.~Tuchin,
  ``Time and space dependence of the electromagnetic field in relativistic heavy-ion collisions,''
  Phys.\ Rev.\ C {\bf 88}, no. 2, 024911 (2013)
  doi:10.1103/PhysRevC.88.024911
  [arXiv:1305.5806 [hep-ph]].
  
\bibitem{Bzdak:2011yy} 
  A.~Bzdak and V.~Skokov,
  ``Event-by-event fluctuations of magnetic and electric fields in heavy ion collisions,''
  Phys.\ Lett.\ B {\bf 710}, 171 (2012).

\bibitem{Tuchin:2014iua} 
  K.~Tuchin,
  ``Electromagnetic field and the chiral magnetic effect in the quark-gluon plasma,''
  Phys.\ Rev.\ C {\bf 91}, no. 6, 064902 (2015)
  doi:10.1103/PhysRevC.91.064902 [arXiv:1411.1363 [hep-ph]].

\bibitem{Li:2016tel} 
  H.~Li, X.~l.~Sheng and Q.~Wang,
  ``Electromagnetic fields with electric and chiral magnetic conductivities in heavy ion collisions,''
  Phys.\ Rev.\ C {\bf 94}, no. 4, 044903 (2016).
  
\bibitem{Tuchin:2010vs} 
  K.~Tuchin,
  ``Synchrotron radiation by fast fermions in heavy-ion collisions,''
  Phys.\ Rev.\ C {\bf 82}, 034904 (2010)
  [Phys.\ Rev.\ C {\bf 83}, 039903 (2011)].
  
\bibitem{Tuchin:2010gx} 
  K.~Tuchin,
  ``Photon decay in strong magnetic field in heavy-ion collisions,''
  Phys.\ Rev.\ C {\bf 83}, 017901 (2011)
  doi:10.1103/PhysRevC.83.017901
  [arXiv:1008.1604 [nucl-th]].
  
\bibitem{Basar:2012bp} 
  G.~Basar, D.~Kharzeev and V.~Skokov,
  ``Conformal anomaly as a source of soft photons in heavy ion collisions,''
  Phys.\ Rev.\ Lett.\  {\bf 109}, 202303 (2012)
  doi:10.1103/PhysRevLett.109.202303
  [arXiv:1206.1334 [hep-ph]].
 
 
\bibitem{Kharzeev:2007jp} 
  D.~E.~Kharzeev, L.~D.~McLerran and H.~J.~Warringa,
  ``The Effects of topological charge change in heavy ion collisions: 'Event by event P and CP violation',''
  Nucl.\ Phys.\ A {\bf 803}, 227 (2008).
 
 
\bibitem{Hirono:2014oda} 
  Y.~Hirono, T.~Hirano and D.~E.~Kharzeev,
  ``The chiral magnetic effect in heavy-ion collisions from event-by-event anomalous hydrodynamics,''
  Phys.\ Rev.\ C {\bf 91}, 054915 (2015).

\bibitem{Son:2012wh} 
  D.~T.~Son and N.~Yamamoto,
  ``Berry Curvature, Triangle Anomalies, and the Chiral Magnetic Effect in Fermi Liquids,''
  Phys.\ Rev.\ Lett.\  {\bf 109}, 181602 (2012).
 
\bibitem{Son:2012zy}D.~T.~Son and N.~Yamamoto,   
``Kinetic theory with Berry curvature from quantum field theories,''   
Phys.\ Rev.\ D {\bf 87}, no. 8, 085016 (2013). 

 \bibitem{Stephanov:2012ki}M.~A.~Stephanov and Y.~Yin,   
``Chiral Kinetic Theory,''   
Phys.\ Rev.\ Lett.\  {\bf 109}, 162001 (2012). 

\bibitem{Gao:2012ix} 
  J.~H.~Gao, Z.~T.~Liang, S.~Pu, Q.~Wang and X.~N.~Wang,
  ``Chiral Anomaly and Local Polarization Effect from Quantum Kinetic Approach,''
  Phys.\ Rev.\ Lett.\  {\bf 109}, 232301 (2012).

\bibitem{Chen:2012ca}J.~W.~Chen, S.~Pu, Q.~Wang and X.~N.~Wang,   
``Berry Curvature and Four-Dimensional Monopoles in the Relativistic Chiral Kinetic Equation,''   
Phys.\ Rev.\ Lett.\  {\bf 110}, no. 26, 262301 (2013). 
 

\bibitem{Hidaka:2016yjf} 
  Y.~Hidaka, S.~Pu and D.~L.~Yang,
  ``Relativistic Chiral Kinetic Theory from Quantum Field Theories,''
  arXiv:1612.04630 [hep-th].

\bibitem{Pu:2014fva}
  S.~Pu, S.~Y.~Wu and D.~L.~Yang,
  ``Chiral Hall Effect and Chiral Electric Waves,''
  Phys.\ Rev.\ D {\bf 91} (2015) no.2,  025011
  doi:10.1103/PhysRevD.91.025011
  [arXiv:1407.3168 [hep-th]].
  
  \bibitem{Chen:2016xtg}
  J.~W.~Chen, T.~Ishii, S.~Pu and N.~Yamamoto,
  ``Nonlinear Chiral Transport Phenomena,''
  arXiv:1603.03620 [hep-th].
  
\bibitem{Ebihara:2017suq} 
  S.~Ebihara, K.~Fukushima and S.~Pu,
  ``Boost invariant formulation of the chiral kinetic theory,''
  arXiv:1705.08611 [hep-ph].
  
  \bibitem{Gorbar:2016qfh}
  E.~V.~Gorbar, I.~A.~Shovkovy, S.~Vilchinskii, I.~Rudenok, A.~Boyarsky and O.~Ruchayskiy,
  ``Anomalous Maxwell equations for inhomogeneous chiral plasma,''
  Phys.\ Rev.\ D {\bf 93} (2016) no.10,  105028
  doi:10.1103/PhysRevD.93.105028
  [arXiv:1603.03442 [hep-th]].


\bibitem{Bhattacharyya:2015pra} 
  A.~Bhattacharyya, S.~K.~Ghosh, R.~Ray and S.~Samanta,
  ``Exploring effects of magnetic field on the Hadron Resonance Gas,''
   arXiv:1504.04533.
 
\bibitem{McInnes:2016dwk} 
  B.~McInnes,
  ``A Rotation/Magnetism Analogy for the Quark Plasma,''
  arXiv:1604.03669 [hep-th].
 
\bibitem{Kharzeev:2013ffa}D.~E.~Kharzeev,   
``The Chiral Magnetic Effect and Anomaly-Induced Transport,''   
Prog.\ Part.\ Nucl.\ Phys.\  {\bf 75}, 133 (2014). 

 \bibitem{Bzdak:2012ia}A.~Bzdak, V.~Koch and J.~Liao,   
``Charge-Dependent Correlations in Relativistic Heavy Ion Collisions and the Chiral Magnetic Effect,''   
Lect.\ Notes Phys.\  {\bf 871}, 503 (2013). 


\bibitem{Kharzeev:2015kna}D.~E.~Kharzeev,   
``Topology, magnetic field, and strongly interacting matter,''   
arXiv:1501.01336 [hep-ph]. 

\bibitem{Tuchin:2013ie} 
  K.~Tuchin,
  ``Particle production in strong electromagnetic fields in relativistic heavy-ion collisions,''
  Adv.\ High Energy Phys.\  {\bf 2013}, 490495 (2013)
  doi:10.1155/2013/490495 [arXiv:1301.0099].
  
\bibitem{Huang:2015oca} 
  X.~G.~Huang,
 ``Electromagnetic fields and anomalous transports in heavy-ion collisions --- A pedagogical review,''
  arXiv:1509.04073 [nucl-th].
   


\bibitem{Shen:2011zc} 
  C.~Shen, S.~A.~Bass, T.~Hirano, P.~Huovinen, Z.~Qiu, H.~Song and U.~Heinz,
  ``The QGP shear viscosity: Elusive goal or just around the corner?,''
  J.\ Phys.\ G {\bf 38}, 124045 (2011).

 \bibitem{Romatschke}P. Romatschke and U. Romatschke, Phys. Rev. Lett.
99, 172301 (2007); M. Luzum and P. Romatschke, Phys. Rev. \textbf{C
78}, 034915 (2008).
   
  %
\bibitem{Heinz}H. Song and U. W. Heinz, 
Phys. Lett. {\bf B 658}, 279 (2008);
Phys. Rev. {\bf C 78}, 024902 (2008).
%

\bibitem{Bozek:2012qs} 
  P.~Bozek and I.~Wyskiel-Piekarska,
  ``Particle spectra in Pb-Pb collisions at $\sqrt{s_{\rm NN}}$  =  2.76 TeV,''
  Phys.\ Rev.\ C {\bf 85}, 064915 (2012)
  doi:10.1103/PhysRevC.85.064915
  [arXiv:1203.6513 [nucl-th]].

\bibitem{Roy:2012jb} 
  V.~Roy, A.~K.~Chaudhuri and B.~Mohanty,
  ``Comparison of results from a 2+1D relativistic viscous hydrodynamic model to elliptic and hexadecapole flow of charged hadrons measured in Au-Au collisions at $\sqrt{s_{\rm {NN}}}$  =  200 GeV,''
  Phys.\ Rev.\ C {\bf 86}, 014902 (2012).
  
\bibitem{Niemi:2012ry} 
  H.~Niemi, G.~S.~Denicol, P.~Huovinen, E.~Molnar and D.~H.~Rischke,
  ``Influence of a temperature-dependent shear viscosity on the azimuthal asymmetries of transverse momentum spectra in ultrarelativistic heavy-ion collisions,''
  Phys.\ Rev.\ C {\bf 86}, 014909 (2012).
  
\bibitem{Heinz:2011kt} 
  U.~Heinz, C.~Shen and H.~Song,
  ``The viscosity of quark-gluon plasma at RHIC and the LHC,''
  AIP Conf.\ Proc.\  {\bf 1441}, 766 (2012).
  
\bibitem{Schenke:2011bn} 
  B.~Schenke, S.~Jeon and C.~Gale,
  ``Higher flow harmonics from (3+1)D event-by-event viscous hydrodynamics,''
  Phys.\ Rev.\ C {\bf 85}, 024901 (2012).



\bibitem{Gursoy:2014aka} 
  U.~Gursoy, D.~Kharzeev and K.~Rajagopal,
  ``Magnetohydrodynamics, charged currents and directed flow in heavy ion collisions,''
  Phys.\ Rev.\ C {\bf 89}, no. 5, 054905 (2014).
  
\bibitem{Zakharov:2014dia} 
  B.~G.~Zakharov,
  ``Electromagnetic response of quark-gluon plasma in heavy-ion collisions,''
  Phys.\ Lett.\ B {\bf 737}, 262 (2014).


\bibitem{Pang:2016yuh} 
  L.~G.~Pang, G.~Endr\"odi and H.~Petersen,
  ``Magnetic-field-induced squeezing effect at energies available at the BNL Relativistic Heavy Ion Collider and at the CERN Large Hadron Collider,''
  Phys.\ Rev.\ C {\bf 93}, no. 4, 044919 (2016)
  doi:10.1103/PhysRevC.93.044919
  [arXiv:1602.06176 [nucl-th]].

\bibitem{Das:2017qfi} 
  A.~Das, S.~S.~Dave, P.~S.~Saumia and A.~M.~Srivastava,
  ``Effects of magnetic field on the plasma evolution in relativistic heavy-ion collisions,''
  arXiv:1703.08162 [hep-ph].

\bibitem{Greif2017} 
  M.~Greif, C.~Greiner and Z.~Xu, ``Magnetic field influence on the early
  time dynamics of heavy-ion collisions,'' arXiv:1704.06505 [hep-ph].


\bibitem{Roy:2015kma} 
  V.~Roy, S.~Pu, L.~Rezzolla and D.~Rischke,
 ``Analytic Bjorken flow in one-dimensional relativistic magnetohydrodynamics,''
  Phys.\ Lett.\ B {\bf 750}, 45 (2015).
 
\bibitem{Pu:2016ayh} 
  S.~Pu, V.~Roy, L.~Rezzolla and D.~H.~Rischke,
  ``Bjorken flow in one-dimensional relativistic magnetohydrodynamics with magnetization,''
  Phys.\ Rev.\ D {\bf 93}, no. 7, 074022 (2016)
  doi:10.1103/PhysRevD.93.074022
  [arXiv:1602.04953 [nucl-th]].
                
  
\bibitem{Pu:2016bxy} 
  S.~Pu and D.~L.~Yang,
  ``Transverse flow induced by inhomogeneous magnetic fields in the Bjorken expansion,''
  Phys.\ Rev.\ D {\bf 93}, no. 5, 054042 (2016)
  doi:10.1103/PhysRevD.93.054042
  [arXiv:1602.04954 [nucl-th]].
  
  
\bibitem{Roy:2015coa} 
  V.~Roy and S.~Pu,
 ``Event-by-event distribution of magnetic field energy over initial fluid energy density in $\sqrt{s_{\rm NN}}$ =  200 GeV Au-Au collisions,''
  Phys.\ Rev.\ C {\bf 92}, 064902 (2015)
  doi:10.1103/PhysRevC.92.064902
  [arXiv:1508.03761 [nucl-th]].
  

\bibitem{Mohapatra:2011ku} 
  R.~K.~Mohapatra, P.~S.~Saumia and A.~M.~Srivastava,
 ``Enhancement of flow anisotropies due to magnetic field in relativistic heavy-ion collisions,''
  Mod.\ Phys.\ Lett.\ A {\bf 26}, 2477 (2011)
  doi:10.1142/S0217732311036711
  [arXiv:1102.3819 [hep-ph]].
  
\bibitem{Tuchin:2011jw} 
  K.~Tuchin,
  ``On viscous flow and azimuthal anisotropy of quark-gluon plasma in strong magnetic field,''
  J.\ Phys.\ G {\bf 39}, 025010 (2012)
  doi:10.1088/0954-3899/39/2/025010
  [arXiv:1108.4394 [nucl-th]].
 
  

\bibitem{Critelli:2014kra} 
  R.~Critelli, S.~I.~Finazzo, M.~Zaniboni and J.~Noronha,
  ``Anisotropic shear viscosity of a strongly coupled non-Abelian plasma from magnetic branes,''
  Phys.\ Rev.\ D {\bf 90}, no. 6, 066006 (2014).


\bibitem{Voronyuk:2011jd} 
  V.~Voronyuk, V.~D.~Toneev, W.~Cassing, E.~L.~Bratkovskaya, V.~P.~Konchakovski and S.~A.~Voloshin,
 ``(Electro-)Magnetic field evolution in relativistic heavy-ion collisions,''
  Phys.\ Rev.\ C {\bf 83}, 054911 (2011).


  
  \bibitem{Rezzolla_book:2013} 
L.~ Rezzolla and O.~ Zanotti, \emph{Relativistic Hydrodynamics},
Oxford University Press,\ Oxford, UK, 2013.

\bibitem{Inghirami:2016iru} 
  G.~Inghirami, L.~Del Zanna, A.~Beraudo, M.~H.~Moghaddam, F.~Becattini
  and M.~Bleicher, ``Numerical magneto-hydrodynamics for relativistic
  nuclear collisions,'' Eur.\ Phys.\ J.\ C {\bf 76}, no. 12, 659 (2016)
  doi:10.1140/epjc/s10052-016-4516-8 [arXiv:1609.03042 [hep-ph]].

\bibitem{degroot}
S.R.~de Groot and L.G.~Suttorp,
\emph{Foundations of electrodynamics},  North-Holland Publishing Company, Amsterdam, 1972.

\bibitem{Israel:1978up} 
  W.~Israel,
  Gen.\ Rel.\ Grav.\  {\bf 9}, 451 (1978).

\bibitem{Kovtun:2016lfw} 
  P.~Kovtun,
  JHEP {\bf 1607}, 028 (2016).
    
\bibitem{Gedalin:1995} 
  M.~Gedalin and I.~Oiberman,
  ``Generally covariant relativistic anisotropic magnetohydrodynamics,''
  Phys.\ Rev.\ E {\bf 51}, 5  (1995).

\bibitem{Huang:2011dc} 
  X.~G.~Huang, A.~Sedrakian and D.~H.~Rischke,
  ``Kubo formulae for relativistic fluids in strong magnetic fields,''
  Annals Phys.\  {\bf 326}, 3075 (2011)
  [arXiv:1108.0602 [astro-ph.HE]].
  
\bibitem{Giacomazzo:2005jy} 
  B.~Giacomazzo and L.~Rezzolla,
  ``The Exact solution of the Riemann problem in relativistic MHD,''
  J.\ Fluid Mech.\  {\bf 562}, 223 (2006)
  [gr-qc/0507102].
  
\bibitem{Bonati:2013lca} 
  C.~Bonati, M.~D'Elia, M.~Mariti, F.~Negro and F.~Sanfilippo,
  ``Magnetic Susceptibility of Strongly Interacting Matter across the Deconfinement Transition,''
  Phys.\ Rev.\ Lett.\  {\bf 111}, 182001 (2013)
  doi:10.1103/PhysRevLett.111.182001
  [arXiv:1307.8063 [hep-lat]].

\bibitem{Huovinen:2009yb} 
  P.~Huovinen and P.~Petreczky,
  ``QCD Equation of State and Hadron Resonance Gas,''
  Nucl.\ Phys.\ A {\bf 837}, 26 (2010)
  doi:10.1016/j.nuclphysa.2010.02.015
  [arXiv:0912.2541 [hep-ph]].
 
\bibitem{Shen:2010uy} 
  C.~Shen, U.~Heinz, P.~Huovinen and H.~Song,
  ``Systematic parameter study of hadron spectra and elliptic flow from viscous hydrodynamic simulations of Au+Au collisions at $\sqrt{s_{NN}} = 200$ GeV,''
  Phys.\ Rev.\ C {\bf 82}, 054904 (2010).

\bibitem{Kolb:1999it} 
  P.~F.~Kolb, J.~Sollfrank and U.~W.~Heinz,
 ``Anisotropic flow from AGS to LHC energies,''
  Phys.\ Lett.\ B {\bf 459}, 667 (1999)
  doi:10.1016/S0370-2693(99)00720-0 [nucl-th/9906003].
  
\bibitem{Roy:2011xt} 
  V.~Roy and A.~K.~Chaudhuri,
 ``Charged particle's elliptic flow in 2+1D viscous hydrodynamics at LHC ($\sqrt{s}$ =  2.76 TeV) energy in Pb+Pb collision,''
  Phys.\ Lett.\ B {\bf 703}, 313 (2011)
  doi:10.1016/j.physletb.2011.08.006
  [arXiv:1103.2870 [nucl-th]].
   
  
\bibitem{Rischke:1995ir} 
  D.~H.~Rischke, S.~Bernard and J.~A.~Maruhn,
  ``Relativistic hydrodynamics for heavy ion collisions. 1. General aspects and expansion into vacuum,''
  Nucl.\ Phys.\ A {\bf 595}, 346 (1995)
  doi:10.1016/0375-9474(95)00355-1 [nucl-th/9504018].
  
  
 
\bibitem{Gupta:2003zh} 
  S.~Gupta,
  ``The Electrical conductivity and soft photon emissivity of the QCD plasma,''
  Phys.\ Lett.\ B {\bf 597}, 57 (2004) [hep-lat/0301006].
  
  
\bibitem{Greif:2014oia} 
  M.~Greif, I.~Bouras, C.~Greiner and Z.~Xu,
  ``Electric conductivity of the quark-gluon plasma investigated using a perturbative QCD based parton cascade,''
  Phys.\ Rev.\ D {\bf 90}, no. 9, 094014 (2014).
  
\bibitem{Finazzo:2013efa} 
  S.~I.~Finazzo and J.~Noronha,
  ``Holographic calculation of the electric conductivity of the strongly coupled quark-gluon plasma near the deconfinement transition,''
  Phys.\ Rev.\ D {\bf 89}, no. 10, 106008 (2014).
 
\bibitem{Bloczynski:2012en} 
  J.~Bloczynski, X.~G.~Huang, X.~Zhang and J.~Liao,
  ``Azimuthally fluctuating magnetic field and its impacts on observables in heavy-ion collisions,''
  Phys.\ Lett.\ B {\bf 718}, 1529 (2013).

  
\bibitem{Bloczynski:2013mca} 
  J.~Bloczynski, X.~G.~Huang, X.~Zhang and J.~Liao,
  ``Charge-dependent azimuthal correlations from AuAu to UU collisions,''
  Nucl.\ Phys.\ A {\bf 939}, 85 (2015).

\bibitem{Deng:2014uja} 
  W.~T.~Deng and X.~G.~Huang,
  ``Electric fields and chiral magnetic effect in Cu+Au collisions,''
  Phys.\ Lett.\ B {\bf 742}, 296 (2015).
  
\bibitem{Deng:2016knn} 
  W.~T.~Deng, X.~G.~Huang, G.~L.~Ma and G.~Wang,
  Phys.\ Rev.\ C {\bf 94}, 041901 (2016).

  

  
  
\end{thebibliography}
\end{document}